\documentclass[preprint,prd,aps,showpacs,showkeys,nofootinbib]{revtex4}
\usepackage{graphicx}
\begin{document}
\title{The naturalness in the BLMSSM and B-LSSM}

\author{Xing-Xing Dong$^{1,2}$\footnote{dxx$\_$0304@163.com},Tai-Fu Feng$^{1,2,3}$\footnote{fengtf@hbu.edu.cn}Shu-Min Zhao$^{1,2}$\footnote{zhaosm@hbu.edu.cn},Hai-Bin Zhang$^{1,2}$\footnote{hbzhang@hbu.edu.cn}}

\affiliation{$^1$ Department of Physics, Hebei University, Baoding 071002, China\\
$^2$ Key Laboratory of High-precision Computation and Application of Quantum Field Theory of Hebei Province, China\\
$^3$ Department of Physics, Chongqing University, Chongqing 401331, China}

\begin{abstract}
In order to interpret the Higgs mass and its decays more naturally, we hope to intrude the BLMSSM and B-LSSM. In the both models, the right-handed neutrino superfields are introduced to better explain the neutrino mass problems. In addition, there are other superfields considered to make these models more natural than MSSM. In this paper, the method of $\chi^2$ analyses will be adopted in the BLMSSM and B-LSSM to calculate the Higgs mass, Higgs decays and muon $g-2$. With the fine-tuning in the region $0.67\%-2.5\%$ and $0.67\%-5\%$, we can obtain the reasonable theoretical values that are in accordance with the experimental results respectively in the BLMSSM and B-LSSM. Meanwhile, the best-fitted benchmark points in the BLMSSM and B-LSSM will be acquired at minimal $(\chi^{BL}_{min})^2 = 2.34736$ and $(\chi^{B-L}_{min})^2 = 2.47754$, respectively.
\end{abstract}


\maketitle
\section{introduction}
The standard model (SM) has been confirmed by many experiments. Especially, the Large Hadron Collider (LHC) have announced a 125.10 GeV SM-like Higgs boson\cite{h0ATLAS,h0CMS,PDG2018}, whose discovery is a great triumph for the SM. Up to now, most of measurements are compatible with the SM predictions at $1\sim2\sigma$ level. More than this, there are still some problems that can not be naturally explained by SM, such as the masses of neutrinos\cite{neutrino1,neutrino2,neutrino3,neutrino4,neutrino5}, the hierarchy problem\cite{hierarchy problem}, the dark matter(DM) candidates\cite{DM2,DM4}, flavor physics\cite{CLFV1,CLFV2} and CP-violating problems\cite{CP problem}.... Therefore, it is necessary to extend SM, and it happens that Minimal Supersymmetric SM (MSSM) is a highly motivated one\cite{MSSM1,MSSM2,MSSM3,MSSM4}.

However, there are still very strong restrictions on supersymmetric parameter space, which will be further explained by the following implications. As we know, the mass of the physical Higgs boson in the MSSM at tree level is less than the Z boson mass , and it can be lifted by the top quark-stop quark loop corrections\cite{stop mass1,stop mass2,stop mass3,stop mass4,stop mass5}. So we need to acquire a rather large stop masses (around TeV region) to give such a large contribution. However, the Higgs soft mass square is deduced as $m_{H_u}^2\simeq-\frac{3y_t^2}{4\pi^2}m_{\tilde{t}}^2\ln\frac{\Lambda}{m_{\tilde{t}}}\sim m_{\tilde{t}}^2$(here $\Lambda$ representing the corresponding new physics(NP) scale while $m_{\tilde{t}}$ corresponding to the scale of the stop mass), and the light stops are good to reproduce the correct scale for electroweak symmetry breaking\cite{stop mass4,stop mass5,stop mass6,stop mass7,stop mass8}. Therefore, we need to introduce the fine-tuning to obtain relatively light stop mass, which can be easily accommodated by introducing an additional contribution to the Higgs
boson mass. Actually, we hope to explain the above problem naturally by extending the MSSM(EMSSM). So far, physicists have proposed many feasible new physical models and in this paper we mainly study the BLMSSM\cite{BLMSSM1,BLMSSM2,BLMSSM3,BLMSSM4,BLMSSM5,BLMSSM6} and B-LSSM\cite{B-LSSM1,B-LSSM2,B-LSSM3,B-LSSM4,B-LSSM5}.

The reason why we discuss the BLMSSM is that the baryon(B) and lepton(L) numbers are local gauge symmetries spontaneously broken at the TeV scale. Not only that, broken baryon number can naturally explain the origin of the matter-antimatter asymmetry in the universe. While broken lepton number can explain the neutrino oscillation experiment well by heavy majorana neutrinos contained in the seesaw mechanism inducing the tiny neutrino masses\cite{BLMSSM1,BLMSSM2,BLMSSM3,BLMSSM4,BLMSSM5}. Additionally, there is a natural suppression of flavour violation in the quark and leptonic sectors since the gauge symmetries and particle content forbid tree level flavor changing neutral currents involving the quarks or charged leptons\cite{BLMSSM1,BLMSSM2,BLMSSM4,BLMSSM5}. Other than this, the mass of the physical Higgs boson can be large without assuming a large stop mass\cite{BLMSSM5}.

Meanwhile, we also study the B-LSSM where gauge symmetry group $SU(3)_C\otimes{SU(2)_L}\otimes{U(1)_Y}\otimes{U(1)_{B-L}}$ is introduced with $B$ representing baryon number and $L$ standing for lepton number. Besides, the invariance under $U(1)_{B-L}$ gauge group imposes the R-parity conservation which is assumed in the MSSM to avoid proton decay\cite{B-L R Parity}. In the B-LSSM, right-handed neutrinos can naturally be implemented due to the introduction of the right-handed neutrino superfields, which can realize type I seesaw mechanism, thus provide an elegant solution for the existence and smallness of the light left-handed neutrino masses. Furthermore, additional parameter space in the B-LSSM is released from the LEP, Tevatron and LHC constraints through the additional singlet Higgs state and right-handed (s)neutrinos to alleviate the hierarchy problem of the MSSM\cite{B-L hierarchy}. Other than this, the model can also provide much more DM candidates comparing that in the MSSM\cite{B-LDM1,B-LDM2,B-LDM3,B-LDM4}.

In this paper, we shall study the natural and realistic EMSSM including both the BLMSSM and B-LSSM by studying the Higgs masses, Higgs decays and muon anomalous magnetic dipole moment(MDM). We first introduce the naturalness conditions specifically in the EMSSM in section II. And the corresponding characteristics for BLMSSM and B-LSSM will be further illustrated in section III. Meanwhile, we derive the concrete theoretical expressions of Higgs decays and muon MDM in both BLMSSM and B-LSSM in section IV. Considering the $\chi^2$ analyses, the numerical results are discussed in section V to satisfy the phenomenological constraints and the relevant experimental data. Last but not least, we summarize the conclusion in section VI. In appendix A, B and C, we give out the corresponding form factors and couplings used in this paper.
\section{naturalness criteria in the EMSSM}
As mentioned in Refs\cite{hierarchy problem,fine-tuning1}, authors popularized a prescription to quantify fine-tuning by an atypical quantity $M_Z$. That is measuring sensitivity in the $Z$ boson mass to general parameters $a_i$ by
\begin{eqnarray}
\Delta_{FT}={\rm Max}\Big\{\Big|\frac{\partial\ln(M_Z^2)}{\partial\ln(a_i)}\Big|\Big\},
\label{DeltaFT}
\end{eqnarray}
here, $a_i$ control the masses of the various supersymmetric partners of the standard particles. The reason for $\Delta_{FT}$ taking maximum is that supersymmetry is responsible for stabilizing the weak scale.

In general weak scale supersymmetric theories, the fine-tuning will be introduced more detail in the Higgs potential. In the MSSM, the SM Higgs-like particle $h^0$ is a linear combination of $H_u$ and $H_d$. The Higgs potential for $h^0$ can be reduced as $V=\bar{m}^2_{h^0}|h^0|^2+\frac{\lambda_{h^0}}{4}|h^0|^4$, where $\bar{m}^2_{h^0}$ is negative and $\lambda_{h^0}$ is positive. Minimizing the Higgs potential, we get $v^2\equiv \langle h^0\rangle^2=-2\bar{m}^2_{h^0}/\lambda_{h^0}$. Then the physical Higgs boson mass can be deduced as $m_{h^0}=-2\bar{m}^2_{h^0}$. So the fine-tuning measure can also be defined as \cite{stop mass4,fine-tuning2,fine-tuning3}
\begin{eqnarray}
\Delta_{FT}\equiv \frac{2\delta\bar{m}^2_{h^0}}{m_{h^0}^2}.
\end{eqnarray}

In general, $\tan \beta\geq2$, so $\bar{m}^2_{h^0}$ can be given as $\bar{m}^2_{h^0}\simeq |\mu|^2+H_u^2|_{tree}+H_u^2|_{rad}$, where $\mu$ is the supersymmetric mass between $H_u$ and $H_d$. $H_u^2|_{tree}$ represent the tree-level contributions to the soft supersymmetry breaking mass square for $H_u$, while $H_u^2|_{rad}$ represent radiative ones. Therefore, we obtain the following concrete bounds
\begin{eqnarray}
\mu\lesssim400 {\rm GeV}\Big(\frac{m_{h^0}}{125.1{\rm GeV}}\Big)\Big(\frac{\Delta_{FT}^{-1}}{5\%}\Big)^{-1/2}.
\end{eqnarray}
Thus, the value of $\mu$ should be smaller than 400 GeV for $5\%$ fine-tuning. Consequently,
the Higgsinos must be light due to the small $\mu$. The dominant contributions to $H_u^2|_{rad}$ arise from stop loop
\begin{eqnarray}
\delta m_{H_u}^2|_{stop}\simeq-\frac{3}{8\pi^2}y_t^2\Big(m_{\tilde{Q}_3}^2+m_{\tilde{U}_3^c}^2+|A_t|^2\Big)\ln\Big(\frac{\Lambda}{m_{\tilde{t}}}\Big),
\end{eqnarray}
where $y_t$ is top quark Yukawa coupling, $m_{\tilde{Q}_3}^2$ $m_{\tilde{U}_3}^2$ and $A_t$ represent the corresponding soft parameters, which determine the stop mass $m_{\tilde{t}}$. Supposing $m_{\tilde{Q}_3}\simeq m_{\tilde{t}_1}$ and $m_{\tilde{U}_3}\simeq m_{\tilde{t}_2}$, we summarize the concrete bound for $M_{\tilde{t}}\equiv\sqrt{m_{\tilde{t}_1}^2+m_{\tilde{t}_2}^2}$
\begin{eqnarray}
&&M_{\tilde{t}}\equiv\sqrt{m_{\tilde{t}_1}^2+m_{\tilde{t}_2}^2}\lesssim\frac{4\pi^2}{3y_t^2}\frac{\Delta_{FT}m_{h^0}^2}{(1+x_t^2)\ln\Big(\Lambda/m_{\tilde{t}}\Big)}
\nonumber\\&&\approx1.2 {\rm TeV}\frac{\sin\beta}{(1+x_t^2)^{1/2}}\Big(\frac{\ln(\Lambda/{\rm TeV})}{3}\Big)^{-1/2}\Big(\frac{m_{h^0}}{125.1{\rm GeV}}\Big)\Big(\frac{\Delta_{FT}^{-1}}{5\%}\Big)^{-1/2},
\end{eqnarray}
where $\tilde{t}_1$ and $\tilde{t}_2$ are stop mass eigenstates and satisfy $\sqrt{m_{\tilde{t}_1}^2+m_{\tilde{t}_2}^2} = A_t/x_t$, Therefore,
we obtain
$M_{\tilde{t}} \lesssim1.2$ TeV for $5\%$ fine-tuning.

Above all, the natural EMSSM should possess relatively small (effective) $\mu$
term as well as stop masses. In this paper, we shall consider the following natural
supersymmetry conditions:

1. The $\mu$ term or effective $\mu$ term is smaller than $400\sqrt{\Delta_{FT}5\%}\;{\rm GeV}$.

2. The square root $M_{\tilde{t}}$ is smaller
than $1.2\sin\beta\sqrt{\Delta_{FT}5\%}\;{\rm TeV}$.
\section{the BLMSSM and B-LSSM}
\subsection{the BLMSSM}
Extending the local gauge group of the SM to $SU(3)_C\otimes{SU(2)_L}\otimes{U(1)_Y}\otimes{U(1)_B}\otimes{U(1)_L}$~\cite{BLMSSM1,BLMSSM4,BLMSSM5}, we obtain a supersymmetric model where baryon $(B)$ and lepton $(L)$ numbers are local gauge symmetries spontaneously broken at the TeV scale(BLMSSM). In order to cancel the $B$ and $L$ anomalies, vector-like families are needed, which are $\hat{Q}_4, \hat{U}_4^c, \hat{D}_4^c, \hat{L}_4, \hat{E}_4^c, \hat{N}_4^c$ and $\hat{Q}_5^c, \hat{U}_5, \hat{D}_5, \hat{L}_5^c, \hat{E}_5, \hat{N}_5 $. Correspondingly, Higgs superfields $\hat{\Phi}_B$ and $\hat{\varphi}_B$ acquire nonzero vacuum expectation values (VEVs) to break baryon number spontaneously, as well as $\hat{\Phi}_L$ and $\hat{\varphi}_L$ are introduced to break lepton number spontaneously. Other than this, in order to make exotic quarks unstable, the model also introduces superfields $\hat{X}$ and $\hat{X}'$. $\hat{X}$ and $\hat{X}'$ mix together, and the lightest mass eigenstate can be a DM candidate.

The superpotential of the BLMSSM is given by\cite{BLsuperpotential}
\begin{eqnarray}
&&{\cal W}_{BL}={\cal W}_{MSSM}+{\cal W}_{B}+{\cal W}_{L}+{\cal W}_{X}\;,
\nonumber\\&&{\cal W}_{B}=\lambda_{Q}\hat{Q}_{4}\hat{Q}_{5}^c\hat{\Phi}_{B}+\lambda_{U}\hat{U}_{4}^c\hat{U}_{5}
\hat{\varphi}_{B}+\lambda_{D}\hat{D}_{4}^c\hat{D}_{5}\hat{\varphi}_{B}+\mu_{B}\hat{\Phi}_{B}\hat{\varphi}_{B}
\nonumber\\&&\hspace{1.2cm}+Y_{{u_4}}\hat{Q}_{4}\hat{H}_{u}\hat{U}_{4}^c+Y_{{d_4}}\hat{Q}_{4}\hat{H}_{d}\hat{D}_{4}^c
+Y_{{u_5}}\hat{Q}_{5}^c\hat{H}_{d}\hat{U}_{5}+Y_{{d_5}}\hat{Q}_{5}^c\hat{H}_{u}\hat{D}_{5},
\nonumber\\&&{\cal W}_{L}=Y_{{e_4}}\hat{L}_{4}\hat{H}_{d}\hat{E}_{4}^c+Y_{{\nu_4}}\hat{L}_{4}\hat{H}_{u}\hat{N}_{4}^c
+Y_{{e_5}}\hat{L}_{5}^c\hat{H}_{u}\hat{E}_{5}+Y_{{\nu_5}}\hat{L}_{5}^c\hat{H}_{d}\hat{N}_{5}
\nonumber\\&&\hspace{1.2cm}+Y_{\nu}\hat{L}\hat{H}_{u}\hat{N}^c+\lambda_{{N^c}}\hat{N}^c\hat{N}^c\hat{\varphi}_{L}
+\mu_{L}\hat{\Phi}_{L}\hat{\varphi}_{L},
\nonumber\\&&{\cal W}_{X}=\lambda_1\hat{Q}\hat{Q}_{5}^c\hat{X}+\lambda_2\hat{U}^c\hat{U}_{5}\hat{X}^\prime
+\lambda_3\hat{D}^c\hat{D}_{5}\hat{X}^\prime+\mu_{X}\hat{X}\hat{X}^\prime.
\end{eqnarray}
where ${\cal W}_{MSSM}$ represents the MSSM superpotential. $\lambda_{Q},\lambda_{U}...$, $Y_{{u_4}},Y_{{d_4}}...$ and $\mu_{B},\mu_{L},\mu_{X}$ are the Yukawa couplings presented in the BLMSSM superpotential. The soft breaking terms in the BLMSSM are generally denoted by\cite{BLsuperpotential,BLsoft}
\begin{eqnarray}
&&{\cal L}_{{soft}}^{BL}={\cal L}_{{soft}}^{MSSM}-(m_{{\tilde{N}^c}}^2)_{{IJ}}\tilde{N}_I^{c*}\tilde{N}_J^c
-m_{{\tilde{Q}_4}}^2\tilde{Q}_{4}^\dagger\tilde{Q}_{4}-m_{{\tilde{U}_4}}^2\tilde{U}_{4}^{c*}\tilde{U}_{4}^c
-m_{{\tilde{D}_4}}^2\tilde{D}_{4}^{c*}\tilde{D}_{4}^c
\nonumber\\
&&\hspace{1.2cm}
-m_{{\tilde{Q}_5}}^2\tilde{Q}_{5}^{c\dagger}\tilde{Q}_{5}^c-m_{{\tilde{U}_5}}^2\tilde{U}_{5}^*\tilde{U}_{5}
-m_{{\tilde{D}_5}}^2\tilde{D}_{5}^*\tilde{D}_{5}-m_{{\tilde{L}_4}}^2\tilde{L}_{4}^\dagger\tilde{L}_{4}
-m_{{\tilde{N}_4}}^2\tilde{N}_{4}^{c*}\tilde{N}_{4}^c
\nonumber\\
&&\hspace{1.2cm}
-m_{{\tilde{E}_4}}^2\tilde{E}_{_4}^{c*}\tilde{E}_{4}^c-m_{{\tilde{L}_5}}^2\tilde{L}_{5}^{c\dagger}\tilde{L}_{5}^c
-m_{{\tilde{N}_5}}^2\tilde{N}_{5}^*\tilde{N}_{5}-m_{{\tilde{E}_5}}^2\tilde{E}_{5}^*\tilde{E}_{5}
-m_{{\Phi_{B}}}^2\Phi_{B}^*\Phi_{B}
\nonumber\\
&&\hspace{1.2cm}
-m_{{\varphi_{B}}}^2\varphi_{B}^*\varphi_{B}-m_{{\Phi_{L}}}^2\Phi_{L}^*\Phi_{L}
-m_{{\varphi_{L}}}^2\varphi_{L}^*\varphi_{L}-\Big(m_{B}\lambda_{B}\lambda_{B}
+m_{L}\lambda_{L}\lambda_{L}+h.c.\Big)
\nonumber\\
&&\hspace{1.2cm}
+\Big\{A_{{u_4}}Y_{{u_4}}\tilde{Q}_{4}H_{u}\tilde{U}_{4}^c+A_{{d_4}}Y_{{d_4}}\tilde{Q}_{4}H_{d}\tilde{D}_{4}^c
+A_{{u_5}}Y_{{u_5}}\tilde{Q}_{5}^cH_{d}\tilde{U}_{5}+A_{{d_5}}Y_{{d_5}}\tilde{Q}_{5}^cH_{u}\tilde{D}_{5}
\nonumber\\
&&\hspace{1.2cm}
+A_{{BQ}}\lambda_{Q}\tilde{Q}_{4}\tilde{Q}_{5}^c\Phi_{B}\hspace{-0.1cm}+\hspace{-0.1cm}A_{{BU}}\lambda_{U}\tilde{U}_{4}^c\tilde{U}_{5}\varphi_{B}
\hspace{-0.1cm}+\hspace{-0.1cm}A_{{BD}}\lambda_{D}\tilde{D}_{4}^c\tilde{D}_{5}\varphi_{B}\hspace{-0.1cm}+\hspace{-0.1cm}B_{B}\mu_{B}\Phi_{B}\varphi_{B}
+h.c.\Big\}
\nonumber\\
&&\hspace{1.2cm}
+\Big\{A_{{e_4}}Y_{{e_4}}\tilde{L}_{4}H_{d}\tilde{E}_{4}^c+A_{{\nu_4}}Y_{{\nu_4}}\tilde{L}_{4}H_{u}\tilde{N}_{4}^c
+A_{{e_5}}Y_{{e_5}}\tilde{L}_{5}^cH_{u}\tilde{E}_{5}+A_{{\nu_5}}Y_{{\nu_5}}\tilde{L}_{5}^cH_{d}\tilde{N}_{5}
\nonumber\\
&&\hspace{1.2cm}
+A_{\nu}Y_{\nu}\tilde{L}H_{u}\tilde{N}^c+A_{{\nu^c}}\lambda_{{\nu^c}}\tilde{N}^c\tilde{N}^c\varphi_{L}
+B_{L}\mu_{L}\Phi_{L}\varphi_{L}+h.c.\Big\}
\nonumber\\
&&\hspace{1.2cm}
+\Big\{A_1\lambda_1\tilde{Q}\tilde{Q}_{5}^cX+A_2\lambda_2\tilde{U}^c\tilde{U}_{5}X^\prime
+A_3\lambda_3\tilde{D}^c\tilde{D}_{5}X^\prime+B_{X}\mu_{X}XX^\prime+h.c.\Big\}\;,
\label{soft-breaking}
\end{eqnarray}
where ${\cal L}_{{soft}}^{MSSM}$ represents the soft breaking terms of the MSSM. Except the squark, slepton and Higgs soft masses $m_{{\tilde{L}_4}}^2,m_{{\tilde{Q}_4}}^2,m_{{\Phi_{B}}}^2...$, there are also other parameters, such as $m_{B},m_{L}...$, $A_{{u_4}},A_{{BQ}}...$, $B_{B},B_{L}...$ and $\tan\beta,\tan\beta_B...$. In our numerical calculation, we adopt the following assumption:
\begin{eqnarray}
&&m_{_{\tilde{Q}_i}}=m_{_{\tilde{U}_i}}=m_{_{\tilde{D}_i}}=m_{_{\tilde{L}_i}}=m_{_{\tilde{R}_i}}
=m_{_{\tilde{N}^c_i}}=m_{_{\tilde{Q}_4}}=m_{_{\tilde{U}_4}}=m_{_{\tilde{D}_4}}=m_{_{\tilde{Q}_5}}\nonumber \\
&&=m_{_{\tilde{U}_5}}=m_{_{\tilde{D}_5}}=m_{_{\tilde{L}_4}}=m_{_{\tilde{N}_4}}=m_{_{\tilde{E}_4}}
=m_{_{\tilde{L}_5}}=m_{_{\tilde{N}_5}}=m_{_{\tilde{E}_5}}\equiv M_0^{BL},\nonumber \\
&&A_l=A_l'=A_u=A_u'=A_d=A_d'=A_{_{u_4}}=A_{_{u_5}}=A_{_{d_4}}=A_{_{d_5}}=A_{_{\nu_4}}=A_{_{e_4}}=A_{_{\nu_5}}\nonumber \\
&&=A_{_{e_5}}=A_{\nu}=A_{{\nu^c}}=A_{_{BQ}}=A_{_{BU}}=A_{_{BD}}\equiv A_0^{BL},m_1=m_2\equiv m_{12}^{BL},g_L=g_B\equiv g_{LB}.
\end{eqnarray}
In the BLMSSM, we mainly consider the effects from parameters $M_0^{BL}$, $A_0^{BL}$, $m_{12}^{BL}$, $g_{LB}^{BL},\mu^{BL}$ and $\tan\beta_{BL}$ for our numerical calculation.
\subsection{the B-LSSM}
In the B-LSSM, one enlarges the local gauge group of the
SM to $SU(3)_C\otimes{SU(2)_L}\otimes{U(1)_Y}\otimes{U(1)_{B-L}}$, where the
${U(1)_{B-L}}$ can be spontaneously broken by the chiral singlet
superfields $\hat{\eta}_1$ and $\hat{\eta}_2$. Besides, the right-handed neutrinos $\hat{\nu}_i^c$ are introduced in the B-LSSM
\begin{eqnarray}
W_{B-L} ={\cal W}_{MSSM}- {\mu'} \hat{\eta}_1\hat{\eta}_2+Y_{x,ij}\hat{\nu}_i^c\hat{\eta}_1\hat{\nu}_j^c+Y_{\nu,ij}\hat{L}_i\hat{H}_2\hat{\nu}_j^c,
\end{eqnarray}
where $i, j$ are generation indices, while $Y_{x,ij}$ and $Y_{\nu,ij}$ are the Yukawa couplings in the B-LSSM superpotential. The soft breaking terms presented in the B-LSSM are written as
\begin{eqnarray}
&&{\cal L}_{{soft}}^{B-L}={\cal L}_{{soft}}^{MSSM}-m_{\tilde{\eta}_1}^2 |\tilde{\eta}_1|^2-m_{\tilde{\eta}_2}^2 |\tilde{\eta}_2|^2
- m_{\tilde{\nu},{i j}}^{2}(\tilde{\nu}^c_{{i}})^* \tilde{\nu}_{{j}}^c+\Big[-{M}_{B B'}\tilde{\lambda}_{B'}\tilde{\lambda}_{B}\nonumber \\
&&\hspace{1.6cm}-\frac{1}{2}{M}_{B'}\tilde{\lambda}_{B'}\tilde{\lambda}_{B'}- B_{\mu'}\tilde{\eta}_1 \tilde{\eta}_2+ T_{\nu}^{i j}H_2 \tilde{\nu}^c_i \tilde{L}_j+T_{x}^{i j}\tilde{\eta}_1 \tilde{\nu}^c_i\tilde{\nu}^c_j+h.c.\Big],
\end{eqnarray}
where $m_{\tilde{\eta}_1}^2 ,m_{\tilde{\eta}_2}^2,m_{\tilde{\nu},{i j}}^{2}...$ are the concrete soft masses. In the B-LSSM, there are also other parameters ${M}_{B B'},M_{B'},B_{\mu'},T_{\nu}^{i j},T_{x}^{i j}...$ and $\tan\beta,\tan\beta'...$. To facilitate numerical discussion, we adopt the following assumption:
\begin{eqnarray}
&&m_{_{{\tilde q,ii}}}=m_{_{{\tilde u,ii}}}=m_{_{{\tilde d,ii}}}=m_{_{{\tilde L,ii}}}=m_{_{{\tilde e,ii}}}
=m_{_{{\tilde\nu,ii}}}\equiv M_0^{B-L},\nonumber \\
&&T_{e,ii}=T_{x,ii}=T_{\nu,ii}=T_{u,ii}=T_{d,ii}\equiv A_0^{B-L},M_1=M_2\equiv m_{12}^{B-L}.
\end{eqnarray}
In the B-LSSM, we mainly consider the effects from parameters $M_0^{B-L}$, $A_0^{B-L}$, $m_{12}^{B-L}$, $g_B^{B-L}$, $g_{YB}^{B-L}$, $\mu^{B-L}$ and $\tan\beta_{B-L}$ for our numerical calculation.
\section{the Higgs decays and $(g-2)_{\mu}$ in the BLMSSM and B-LSSM }
In the EMSSM, we consider the radiative corrections from exotic fermions and corresponding supersymmetric partners to obtain the physical Higgs mass. The corrections to Higgs masses in the BLMSSM were discussed specifically in Ref.\cite{BLsuperpotential}, while the ones in the B-LSSM were introduced concretely in Refs.\cite{B-LSSM5,B-LHiggs1}. The corresponding parameter constraints in the BLMSSM and B-LSSM are considered respectively in this paper. Then the Higgs decays and $(g-2)_{\mu}$ will be taken over explicitly as follows.
\subsection{the Higgs decays}
The LHC produces the Higgs chiefly from the gluon fusion. Meanwhile, the leading order(LO) contributions for $ h^0\rightarrow gg$ originate from the one-loop diagrams, which can be modified through virtual top quark in the SM. In the EMSSM, the LO contributions need to be added by the Higgs-new particle couplings, whose effects are significant. So the decay width of $ h^0\rightarrow gg$ can be shown as\cite{BLsuperpotential,Higgs decay1,Higgs decay2,Higgs decay3,Higgs decay5}
\begin{eqnarray}
&&\Gamma_{{NP}}(h^0\rightarrow gg)={G_{F}\alpha_s^2m_{{h^0}}^3\over64\sqrt{2}\pi^3}
\Big|\sum\limits_{q}g_{{h^0qq}}A_{1/2}(x_q)
+\sum\limits_{\tilde q}g_{{h^0\tilde{q}\tilde{q}}}{m_{{\rm Z}}^2\over m_{{\tilde q}}^2}A_{0}(x_{{\tilde{q}}})\Big|^2\;,
\label{hgg}
\end{eqnarray}
with $x_a=m_{{h^0}}^2/(4m_a^2)$. $q$ and $\tilde{q}$ denote the concrete quarks and squarks in the EMSSM.

The LO contributions for decay $h^0\rightarrow \gamma\gamma$ also originate from one-loop diagrams. In the SM, the concrete contributions are mainly derived from top quark and charged gauge boson $W^{\pm}$. Due to the Higgs-new particle couplings in the EMSSM, the decay width of $h^0\rightarrow \gamma\gamma$ can be expressed as\cite{BLsuperpotential,Higgs decay1,Higgs decay2,Higgs decay3,Higgs decay5,Higgs decay6,Higgs decay7}
\begin{eqnarray}
&&\hspace{-0.8cm}\Gamma_{{NP}}(h^0\rightarrow\gamma\gamma)={G_{F}\alpha^2m_{{h^0}}^3\over128\sqrt{2}\pi^3}
\Big|\sum\limits_fN_cQ_{f}^2g_{{h^0ff}}A_{1/2}(x_f)+\sum\limits_{\tilde f}N_cQ_{f}^2g_{{h^0\tilde{f}\tilde{f}}}{m_{ Z}^2\over m_{{\tilde f}}^2}
A_{0}(x_{{\tilde{f}}})
\nonumber\\&&\hspace{1.5cm}+g_{{h^0H^+H^-}}{m_{{\rm Z}}^2\over m_{{H^\pm}}^2}A_0(x_{{H^\pm}})+g_{{h^0WW}}A_1(x_{{\rm W}})
+\sum\limits_{i=1}^2g_{{h^0\chi_i^+\chi_i^-}}{m_{{\rm W}}\over m_{{\chi_i}}}A_{1/2}(x_{{\chi_i}})
\Big|^2\;.
\label{hpp}
\end{eqnarray}

The decay width for $h^0\rightarrow ZZ, WW$ are given by\cite{Higgs decay8,Higgs decay9}
\begin{eqnarray}
&&\Gamma(h^0\rightarrow WW)={3e^4m_{{h^0}}\over512\pi^3s_{ W}^4}|g_{h^0WW}|^2
F({m_{_{\rm W}}\over m_{h^0}}),\;\nonumber\\
&&\Gamma(h^0\rightarrow ZZ)={e^4m_{{h^0}}\over2048\pi^3s_{W}^4c_{W}^4}|g_{h^0ZZ}|^2
\Big(7-{40\over3}s_{W}^2+{160\over9}s_{W}^4\Big)F({m_{Z}\over m_{_{h^0}}}).
\end{eqnarray}

With the Born approximation, the decay width of the physical Higgs into fermion pairs $h^0\rightarrow f\bar{f}$ is written as\cite{Higgs decay10}
\begin{eqnarray}
&&\Gamma(h^0\rightarrow f\bar{f})=N_c{G_Fm_f^2m_{{h^0}}\over4\sqrt{2}\pi}|g_{h^0ff}|^2
(1-{4m_f^2\over m_{h^0}^2})^{3/2},\;
\end{eqnarray}
where the form factors $A_{1/2}(x)$, $A_0(x)$, $A_1(x)$ and $F(x)$ are summarized in the appendix A. In the BLMSSM, the concrete expressions for $g_{{h^0qq}}$, $g_{{h^0\tilde{q}\tilde{q}}}$, $g_{{h^0ff}}$, $g_{{h^0H^+H^-}}$, $g_{{h^0\chi_i^+\chi_i^-}}$, $g_{{h^0\tilde{f}\tilde{f}}}$, $g_{{h^0WW}}$ and $g_{h^0ZZ}$ have been discussed in Ref.\cite{BLsuperpotential}. The relevant expressions that present in the B-LSSM are specifically discussed in the following appendix B.

The signal strengths for the Higgs decay channels are quantified by the following ratios\cite{Higgs ratio}
\begin{eqnarray}
&&\mu_{\gamma\gamma,VV^*}^{ggF}=\frac{\sigma_{NP}(ggF)}{\sigma_{SM}(ggF)}
\frac{BR_{NP}(h^0\rightarrow \gamma\gamma,VV^*)}{BR_{SM}(h^0\rightarrow \gamma\gamma,VV^*)},(V=Z,W),\nonumber \\&&\mu_{f\bar{f}}^{VBF}=\frac{\sigma_{NP}(VBF)}{\sigma_{SM}(VBF)}
\frac{BR_{NP}(h^0\rightarrow f\bar{f})}{BR_{SM}(h^0\rightarrow f\bar{f})},(f=b,\tau),
\end{eqnarray}
where ggF and VBF stand for gluon-gluon fusion and vector boson fusion respectively. Meanwhile, $\mu_{\gamma\gamma,VV^*}$ are mainly affected by gluon-gluon fusion while $\mu_{f\bar{f}}$ is more likely to be influenced by vector boson fusion. The Higgs production cross sections can be further simplified as
$\frac{\sigma_{NP}(ggF)}{\sigma_{SM}(ggF)}\approx
\frac{\Gamma_{NP}(h^0\rightarrow gg)}{\Gamma_{SM}(h^0\rightarrow gg)}
,\frac{\sigma_{NP}(VBF)}{\sigma_{SM}(VBF)}\hspace{-0.1cm}\approx\hspace{-0.1cm}
\frac{\Gamma_{NP}(h^0\hspace{-0.1cm}\rightarrow\hspace{-0.1cm} VV^*)}{\Gamma_{SM}(h^0\hspace{-0.1cm}\rightarrow\hspace{-0.1cm} VV^*)}$.
Therefore, the ratios of the signal strengths from the Higgs decay channels are reduced as
\begin{eqnarray}
&&\hspace{-0.8cm}\mu_{\gamma\gamma}^{ggF}\approx\frac{\Gamma_{NP}(h^0\rightarrow gg)}{\Gamma_{SM}(h^0\rightarrow gg)}
\frac{\Gamma_{NP}(h^0\rightarrow \gamma\gamma)/\Gamma_{NP}^{h^0}}{\Gamma_{SM}(h^0\rightarrow \gamma\gamma)/\Gamma_{SM}^{h^0}}=\frac{\Gamma_{SM}^{h^0}}{\Gamma_{NP}^{h^0}}\frac{\Gamma_{NP}(h^0\rightarrow gg)}{\Gamma_{SM}(h^0\rightarrow gg)}
\frac{\Gamma_{NP}(h^0\hspace{-0.1cm}\rightarrow\hspace{-0.1cm} \gamma\gamma)}{\Gamma_{SM}(h^0\hspace{-0.1cm}\rightarrow\hspace{-0.1cm} \gamma\gamma)},\nonumber \\
&&\hspace{-0.8cm}\mu_{VV^*}^{ggF}\hspace{-0.1cm}\approx\hspace{-0.1cm}\frac{\Gamma_{NP}(h^0\hspace{-0.1cm}\rightarrow\hspace{-0.1cm} gg)}{\Gamma_{SM}(h^0\hspace{-0.1cm}\rightarrow\hspace{-0.1cm} gg)}
\frac{\Gamma_{NP}(h^0\hspace{-0.1cm}\rightarrow\hspace{-0.1cm} VV^*)/\Gamma_{NP}^{h^0}}{\Gamma_{SM}(h^0\hspace{-0.1cm}\rightarrow\hspace{-0.1cm} VV^*)/\Gamma_{SM}^{h^0}}
\hspace{-0.1cm}=\hspace{-0.1cm}\frac{\Gamma_{SM}^{h^0}}{\Gamma_{NP}^{h^0}}\frac{\Gamma_{NP}(h^0\hspace{-0.1cm}\rightarrow \hspace{-0.1cm}gg)}{\Gamma_{SM}(h^0\hspace{-0.1cm}\rightarrow \hspace{-0.1cm}gg)}|g_{h^0VV}|^2,(V\hspace{-0.1cm}=\hspace{-0.1cm}Z,W),\nonumber \\
&&\hspace{-0.8cm}\mu_{f\bar{f}}^{VBF}\approx\frac{\Gamma_{NP}(h^0\rightarrow VV^*)}{\Gamma_{SM}(h^0\rightarrow VV^*)}
\frac{\Gamma_{NP}(h^0\rightarrow f\bar{f})/\Gamma_{NP}^{h^0}}{\Gamma_{SM}(h^0\rightarrow f\bar{f})/\Gamma_{SM}^{h^0}}=\frac{\Gamma_{SM}^{h^0}}{\Gamma_{NP}^{h^0}}|g_{h^0VV}|^2|g_{h^0ff}|^2,(f=b,\tau),
\end{eqnarray}
here, $\Gamma_{NP}^{h^0}\hspace{-0.1cm}=\hspace{-0.1cm}\sum_{f}\Gamma_{NP}\hspace{-0.1cm}(h^0\hspace{-0.1cm}\rightarrow \hspace{-0.1cm}f\bar{f})
+\sum_{V}\hspace{-0.1cm}\Gamma_{NP}(h^0\hspace{-0.1cm}\rightarrow \hspace{-0.1cm}VV^*)+\Gamma_{NP}(h^0\hspace{-0.1cm}\rightarrow\hspace{-0.1cm} gg)+\Gamma_{NP}(h^0\hspace{-0.1cm}\rightarrow\hspace{-0.1cm} \gamma\gamma)$ represents the NP total decay width of physical Higg, $\frac{\Gamma_{NP}(h^0\rightarrow VV^*)}{\Gamma_{SM}(h^0\rightarrow VV^*)}=|g_{h^0VV}|^2$ and $\frac{\Gamma_{NP}(h^0\rightarrow f\bar{f})}{\Gamma_{SM}(h^0\rightarrow f\bar{f})}=|g_{h^0ff}|^2$.
\subsection{$(g-2)_{\mu}$}
\begin{figure}[t]
\centering
\includegraphics[width=8cm]{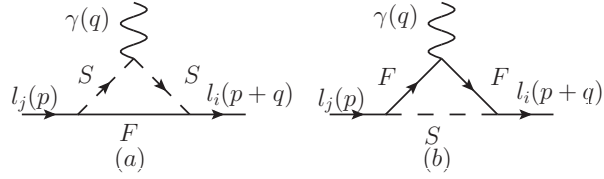}\\
\caption{The one-loop diagrams affect $(g-2)_{\mu}$ in the BLMSSM and B-LSSM.} \label{fig1}
\end{figure}

The effective Lagrangian for the muon MDM can be actually summarized as follows
\begin{eqnarray}
&&{\cal L}_{MDM}={e\over4m_{\mu}}\;a_{\mu}\;\bar{l}_{\mu}\sigma^{\alpha\beta}
l_{\mu}\;F_{{\alpha\beta}},\label{adm}
\end{eqnarray}
where $\sigma_{\alpha\beta}=i[\gamma_\alpha,\gamma_\beta]/2$, $F_{\alpha\beta}$ is the electromagnetic field strength. Other than this, $l_{\mu}$ denotes the muon
fermion, $m_{\mu}$ represents the corresponding muon mass and $a_\mu$ is the muon MDM. Generally, we obtain the muon MDM through the effective Lagrangian method\cite{MSSM4,g-21,g-22}
\begin{eqnarray}
&&a_\mu=\frac{4Q_fm_\mu^2}{(4\pi)^2}\Re(C_2^++C_2^{-*}+C_6^+).
\end{eqnarray}
where $Q_f=-1$, $C_{2,6}^{\pm}$ represent the Wilson coefficients of the corresponding operators $\mathcal{O}_{2,6}^{\mp}$
\begin{eqnarray}
\mathcal{O}_2^{\mp}=\frac{eQ_f}{(4\pi)^2}\overline{(i\mathcal{D}_{\mu}l_{\mu})}\gamma^{\mu}
F\cdot\sigma\omega_{\mp}l_{\mu},\;\;\;
\mathcal{O}_6^{\mp}=\frac{eQ_fm_\mu}{(4\pi)^2}\bar{l}_{\mu}F\cdot\sigma
\omega_{\mp}l_{\mu},
\end{eqnarray}
with $\mathcal{D}_{\mu}=\partial_{\mu}+ieA_{\mu}$ and $\omega_{\mp}=\frac{(1\mp\gamma_5)}{2}$. The EMSSM contributions to muon MDM originate from the one-loop triangle diagrams, which are shown in FIG.\ref{fig1}.
So the one-loop corrections to muon MDM can be expressed as
\begin{eqnarray}
&&\Delta a_\mu=a_\mu(a)+a_\mu(b).
\label{oneloop}
\end{eqnarray}

In the EMSSM, the muon MDM corresponding to FIG. \ref{fig1}(a) can be formulated as
\begin{eqnarray}
&&a_\mu(a)=
-\sum_{F}\sum_{S}\Big[\Re[(\mathcal{A}_1)^I(\mathcal{A}_2)^{I*}]
y_S\sqrt{y_Fy_{m_{\mu}}}\;\frac{\partial^2 \mathcal{B}(y_F,y_S)}{\partial y_S^2}
\nonumber\\&&\hspace{1.4cm}+\frac{1}{3}(|(\mathcal{A}_1)^I|^2+|(\mathcal{A}_2)^I|^2)y_Sy_{m_{\mu}}
\frac{\partial\mathcal{B}_1(y_F,y_S)}{\partial y_S}\Big],
\end{eqnarray}
where $y_i$ denote $\frac{m_i^2}{\Lambda^2}$. $\mathcal{B}(x,y),\;\mathcal{B}_1(x,y)$ are the one-loop functions and given out in appendix A. Similarly, the muon MDM for FIG.\ref{fig1}(b) is deduced as follows
\begin{eqnarray}
&&a_\mu(b)=
\sum_{F}\sum_{S}\Big[-2\Re[(\mathcal{C}_1)^I(\mathcal{C}_2)^{I*}]
\sqrt{y_Fy_{m_{\mu}}}\;\mathcal{B}_1(y_S,y_F)
\nonumber\\&&\hspace{1.4cm}+\frac{1}{3}(|(\mathcal{C}_1)^I|^2+|(\mathcal{C}_2)^I|^2)y_Fy_{m_{\mu}}
\frac{\partial\mathcal{B}_1(y_S,y_F)}{\partial y_F}\Big].
\end{eqnarray}

In the BLMSSM, the concrete expressions for $(\mathcal{A}_1)^I,(\mathcal{A}_2)^I,(\mathcal{C}_1)^I$ and $(\mathcal{C}_2)^I$ can be found in Ref.\cite{g-24}. The corresponding expressions that present in the B-LSSM will be specifically discussed in the following appendix C.
\section{$\chi^2$ analyses for the numerical results}
In this paper, we will consider the $\chi^2$ analyses for the corresponding theoretical and experimental data in both BLMSSM and B-LSSM. In general, the expression for $\chi^2$ can be simplified with $\xi$ data points as\cite{chi1,chi2}
\begin{eqnarray}
\chi^2=\sum_{\xi}(\frac{\mu_{\xi}^{th}-\mu_{\xi}^{exp}}{\delta_{\xi}})^2,
\end{eqnarray}
in which the theoretical values obtained for our model $\mu_{\xi}^{th}$ are confronted with the experimental measurements $\mu_{\xi}^{exp}$, $\delta_{\xi}$ represent the errors which include both statistic and system.

Actually, combining the experimental results from ATLAS, CMS, LHC and TEVA collaborations, we adopt the averages for Higgs decays from PDG\cite{PDG2018}, which are $\mu_{\gamma\gamma}^{exp}=1.10^{+0.10}_{-0.09}$\cite{h0gamma21,h0gamma22,h0gamma23,h0gamma24}, $\mu_{WW}^{exp}=1.08^{+0.18}_{-0.16}$\cite{h0gamma23,h0gamma24}, $\mu_{ZZ}^{exp}=1.19^{+0.12}_{-0.11}$\cite{h0gamma23,h0ZZ1,h0ZZ2}, $\mu_{b\bar{b}}^{exp}=1.02\pm0.15$\cite{h0gamma23,h0gamma24,h0bb1,h0bb2} and $\mu_{\tau\bar{\tau}}^{exp}=1.11 \pm0.17$\cite{h0gamma23,h0gamma24,h0tautau}. Furthermore, the muon MDM possesses $3.7\sigma$ deviation between experimental data and theoretical prediction:
$\Delta a_{\mu} =a^{exp}_{\mu}-a^{SM}_{\mu}=(274\pm73)\times 10^{-11}$\cite{deltaau1,deltaau4}. Considering the constraints $\mu\lesssim400\sqrt{\Delta_{FT}5\%}{\rm GeV}$, $M_{\tilde{t}}\lesssim1.2\sin\beta\sqrt{\Delta_{FT}5\%}{\rm TeV}$ and $115{\rm GeV}\lesssim m_{h^0}\lesssim135{\rm GeV}$, the numerical analyses will be further discussed as follows.
\begin{figure}[t]
\centering
\includegraphics[width=4.cm]{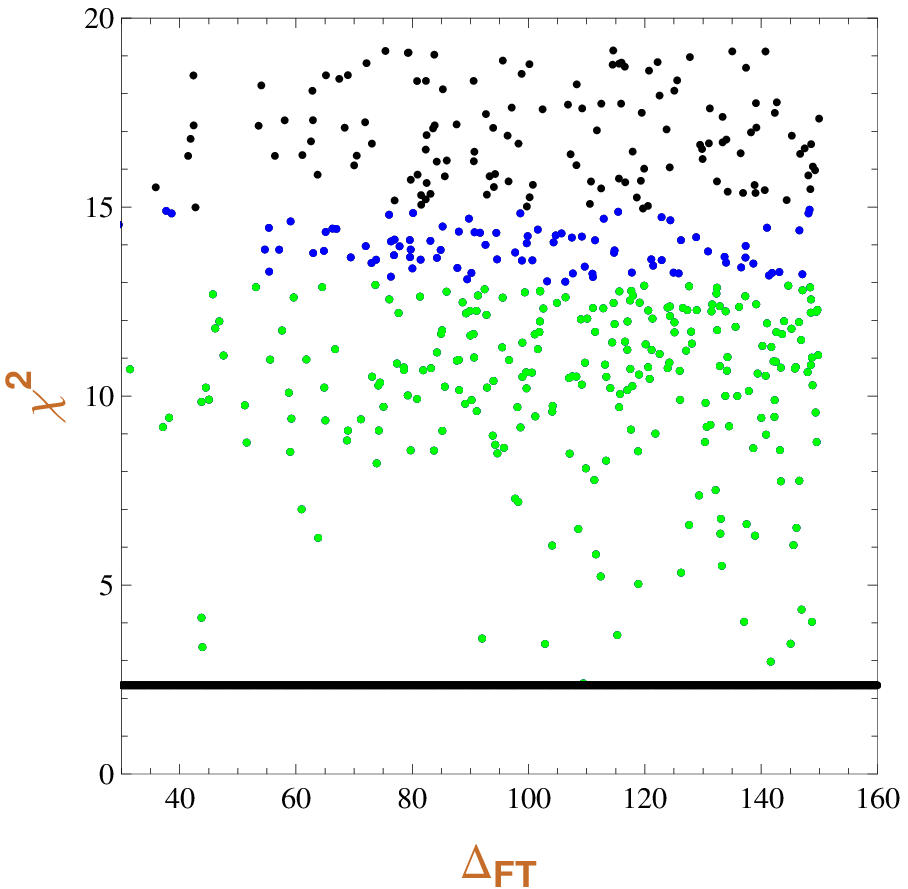}
\includegraphics[width=4.15cm]{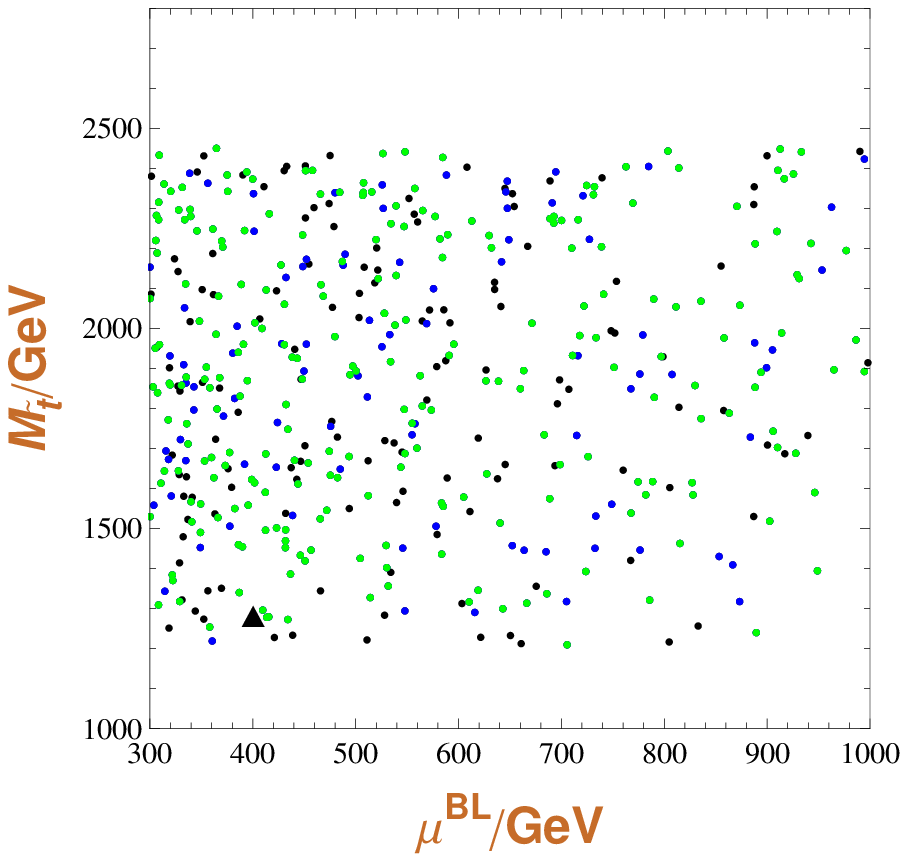}
\includegraphics[width=3.85cm]{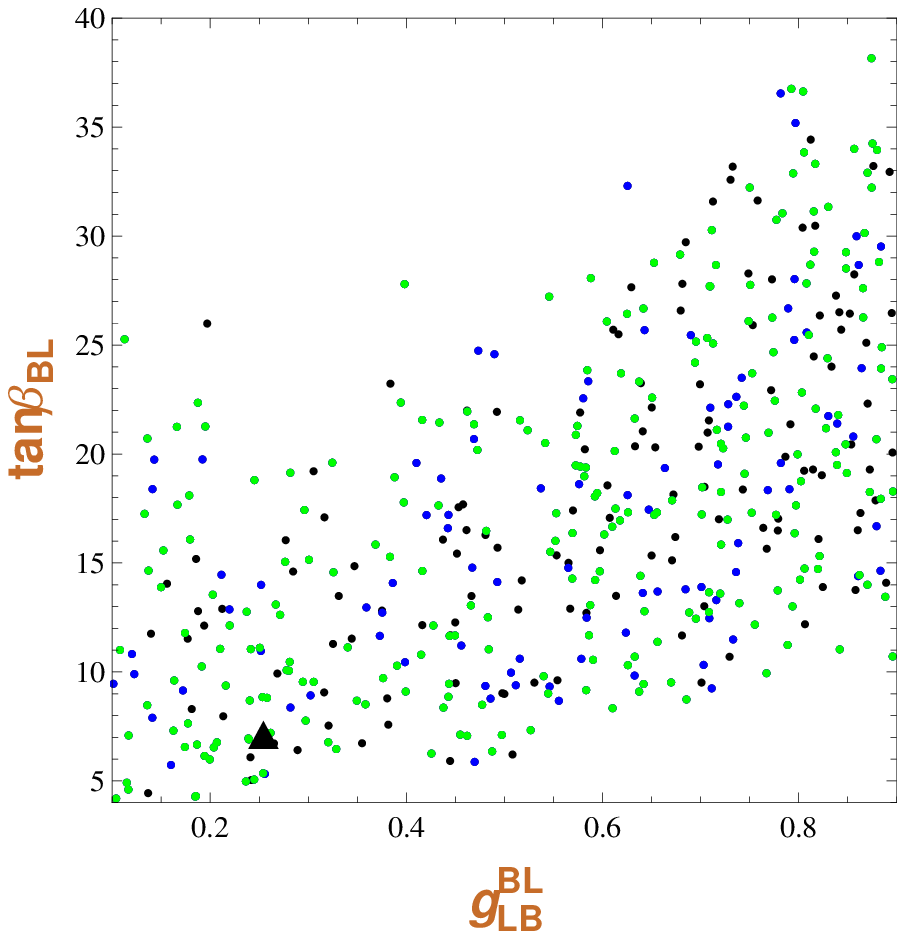}
\includegraphics[width=4.1cm]{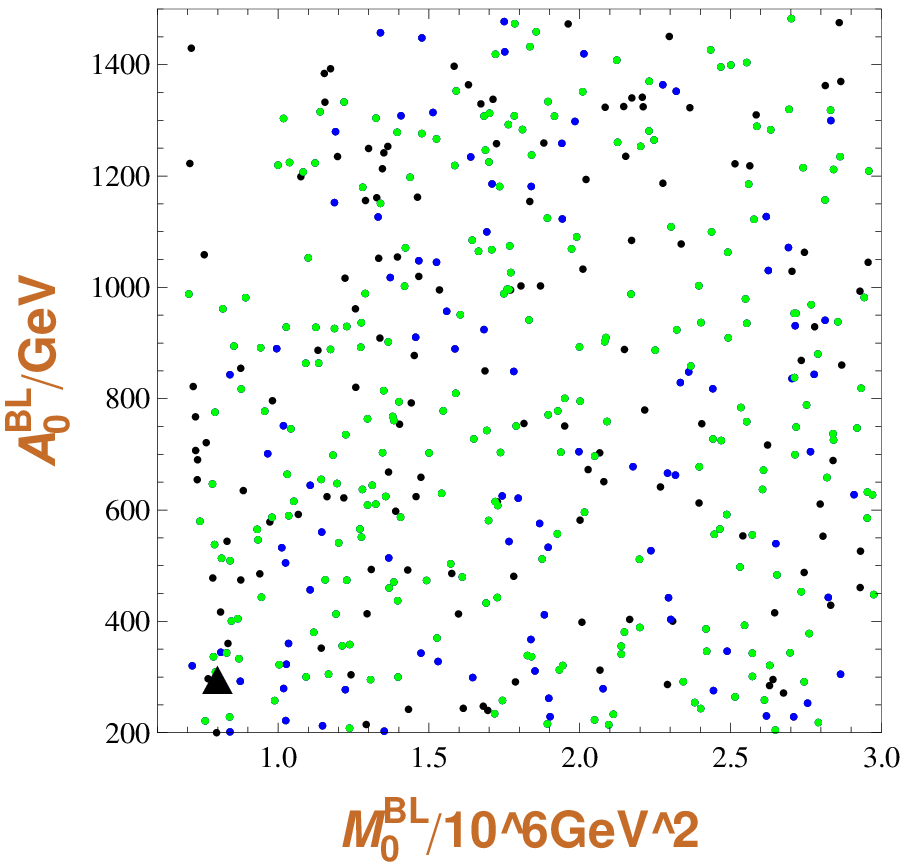}\\
\includegraphics[width=4.1cm]{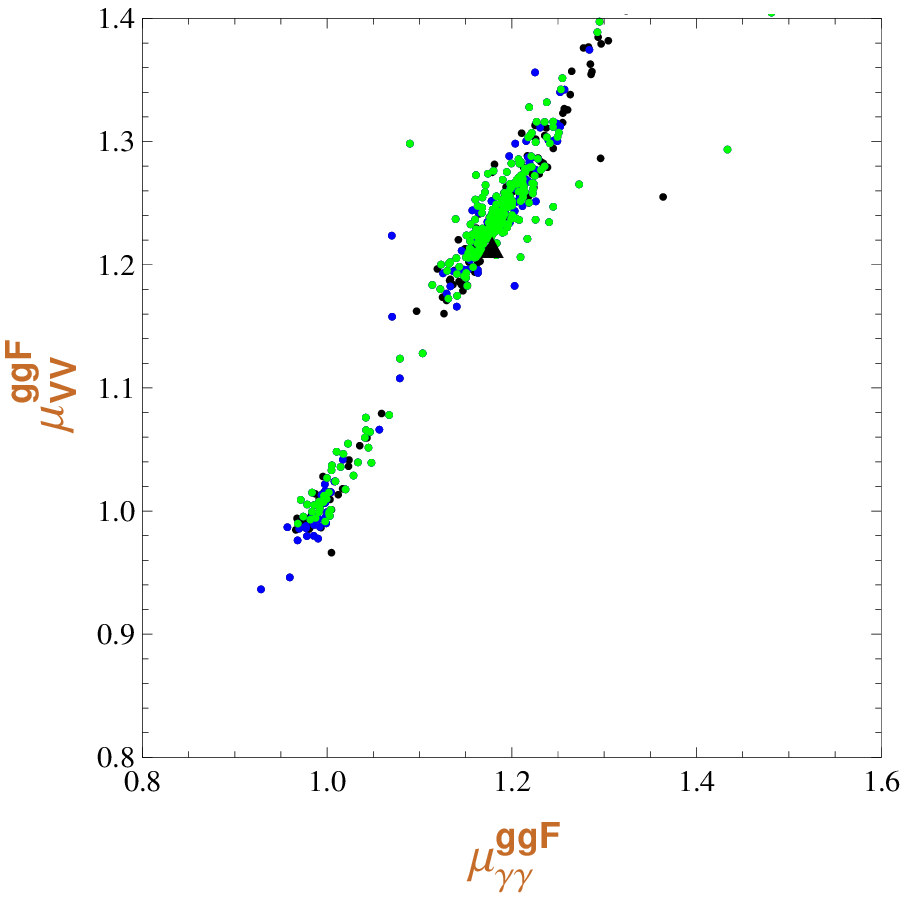}
\includegraphics[width=4.1cm]{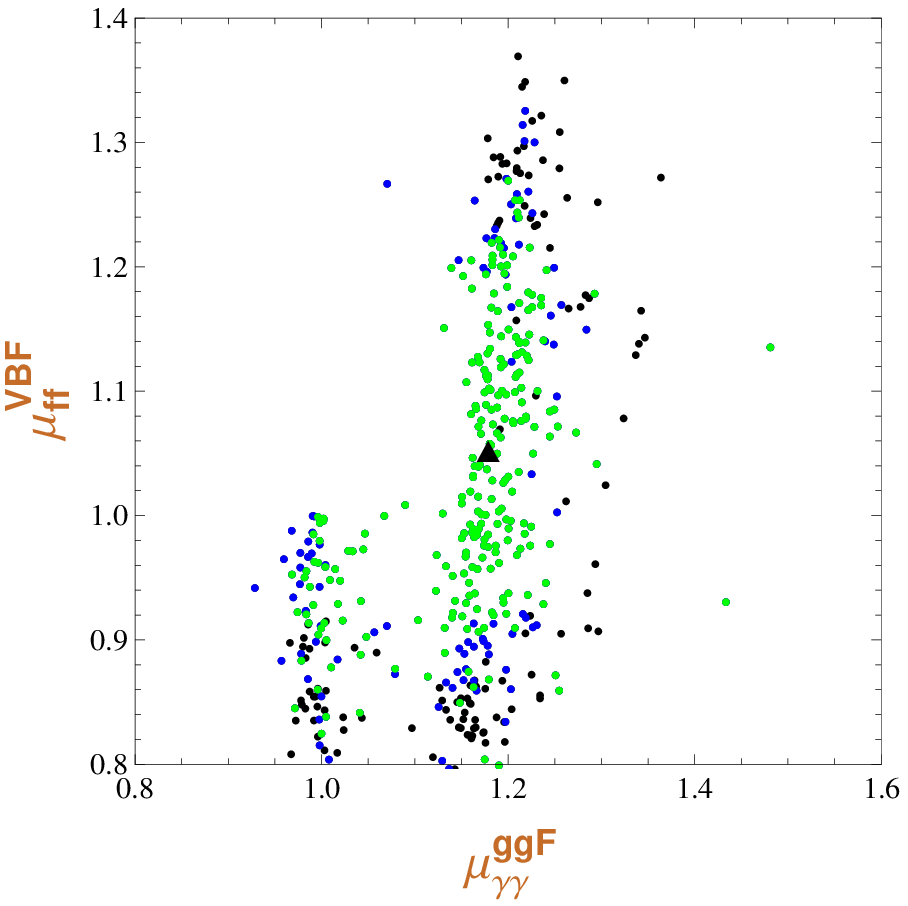}
\includegraphics[width=4.35cm]{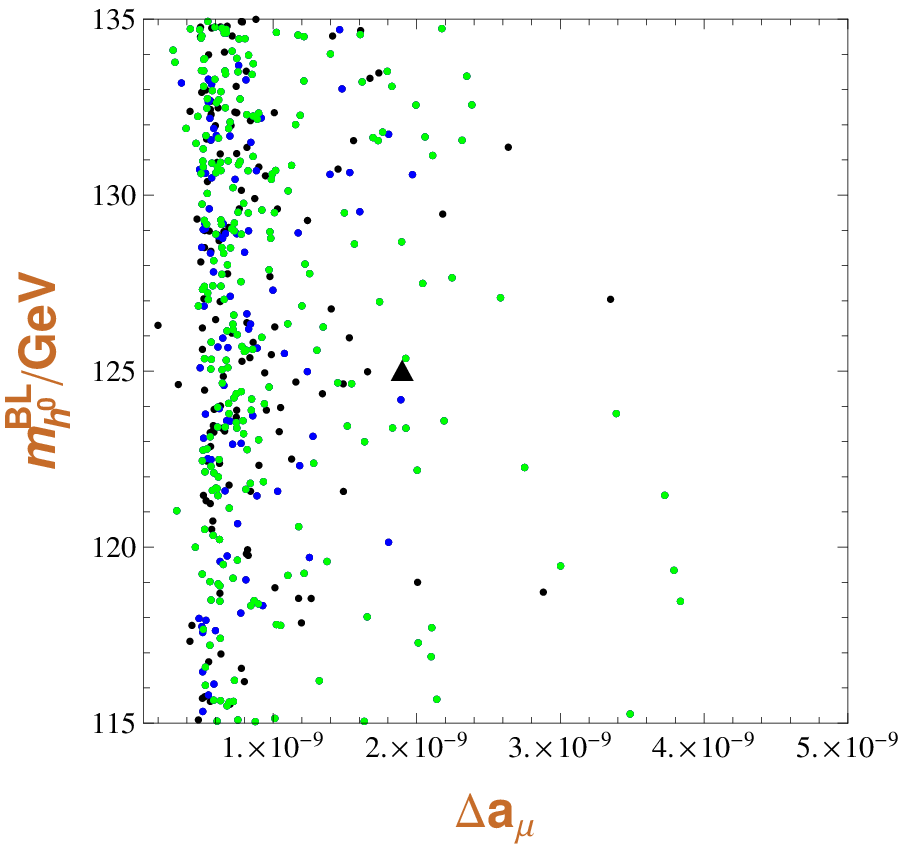}
\caption{The fitting results for BLMSSM with the $\chi^2$ analyses.} \label{fig2}
\end{figure}

First of all, we analyze the numerical results in the BLMSSM. With $\Delta_{FT}$ in the region of $40\sim150$, we propose the $\Delta_{FT}$ versus $\chi^2$, $\mu^{BL}$ versus $M_{\tilde{t}}$, $g_{LB}^{BL}$ versus $\tan\beta_{BL}$, $M_0^{BL}$ versus $A_0^{BL}$, $\mu_{\gamma\gamma}^{ggF}$ versus $\mu_{VV}^{ggF}$, $\mu_{\gamma\gamma}^{ggF}$ versus $\mu_{ff}^{VBF}$and $\Delta a_{\mu}$ versus $m_{h^0}^{BL}$ in FIG.\ref{fig2}. The black triangle shows the best-fitted benchmark point with minimal $(\chi^{BL}_{min})^2$ = 2.34736. The green, blue, and black regions are respectively $90\%$, $95\%$, and $99\%$ confidence levels with $\chi^2 < (\chi^{BL}_{min})^2+ 10.65$, $(\chi^{BL}_{min})^2 + 12.59$ and $(\chi^{BL}_{min})^2+ 16.81$. It is clear to see that $g_{LB}^{BL}$ is changing from 0.1 to 0.9 while $\tan\beta_{BL}$ is in the region $4\sim40$. Not only that, $\mu_{\gamma\gamma}^{ggF}$ and $\mu_{VV}^{ggF}$ are both around $1.0\sim1.3$ and $\mu_{ff}^{VBF}$ is fixed in the range of 0.9 to 1.2, whose parameter spaces for $\chi^2 < (\chi^{BL}_{min})^2+ 10.65$ are obviously smaller than that for $\chi^2 <(\chi^{BL}_{min})^2 + 12.59$ and $(\chi^{BL}_{min})^2+ 16.81$. $\Delta a_{\mu}$ can be limited to $1.0\times10^{-9}\sim3.0\times10^{-9}$ with the fine-tuning in the region $0.67\%-2.5\%$. So $\mu_{\gamma\gamma}^{ggF}$, $\mu_{VV}^{ggF}$, $\mu_{VV}^{ggF}$ and $\Delta a_{\mu}$ which agree well with the concrete experimental results can naturally be explained in the BLMSSM.
\begin{figure}[t]
\centering
\includegraphics[width=5cm]{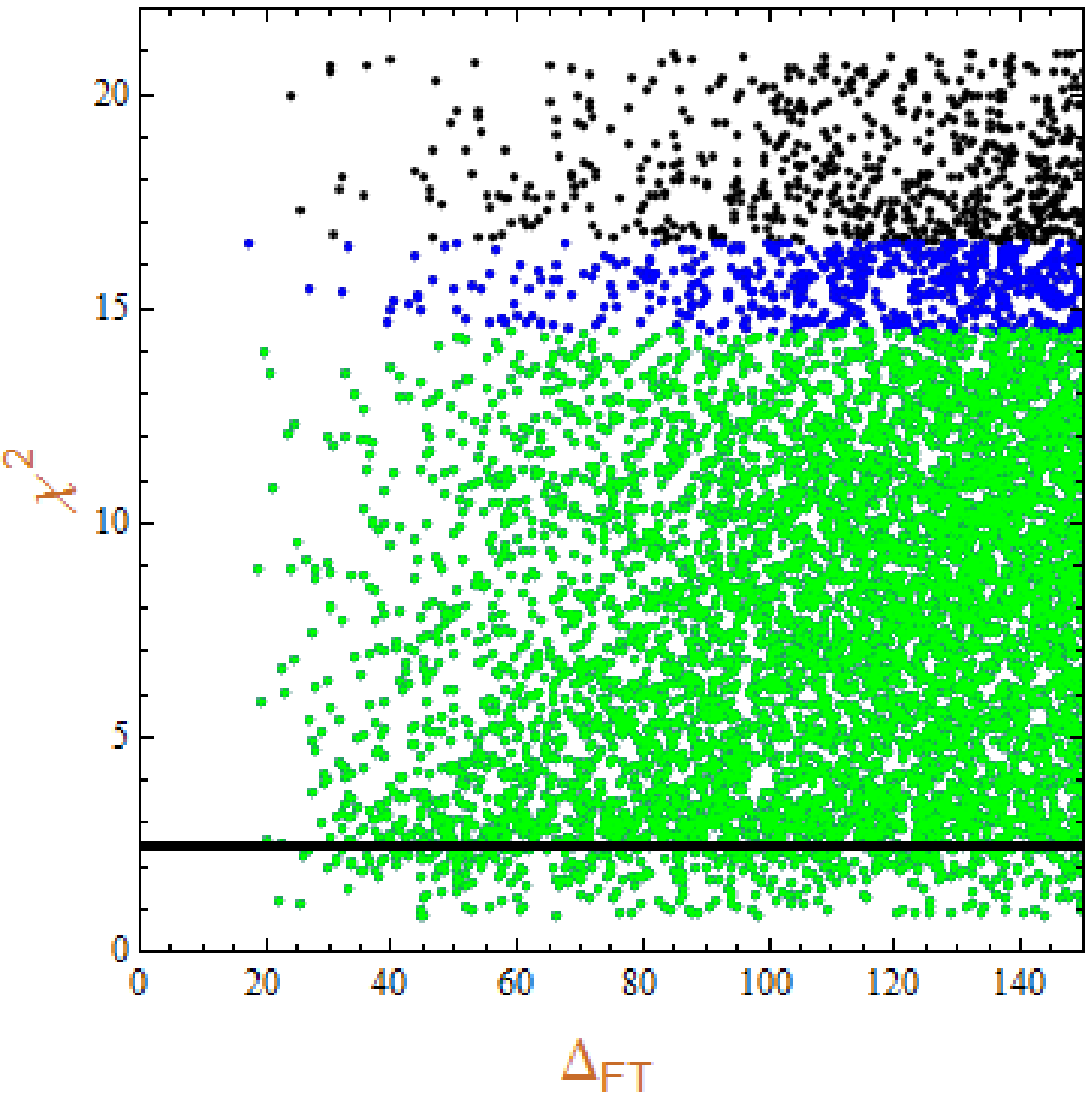}
\includegraphics[width=5.35cm]{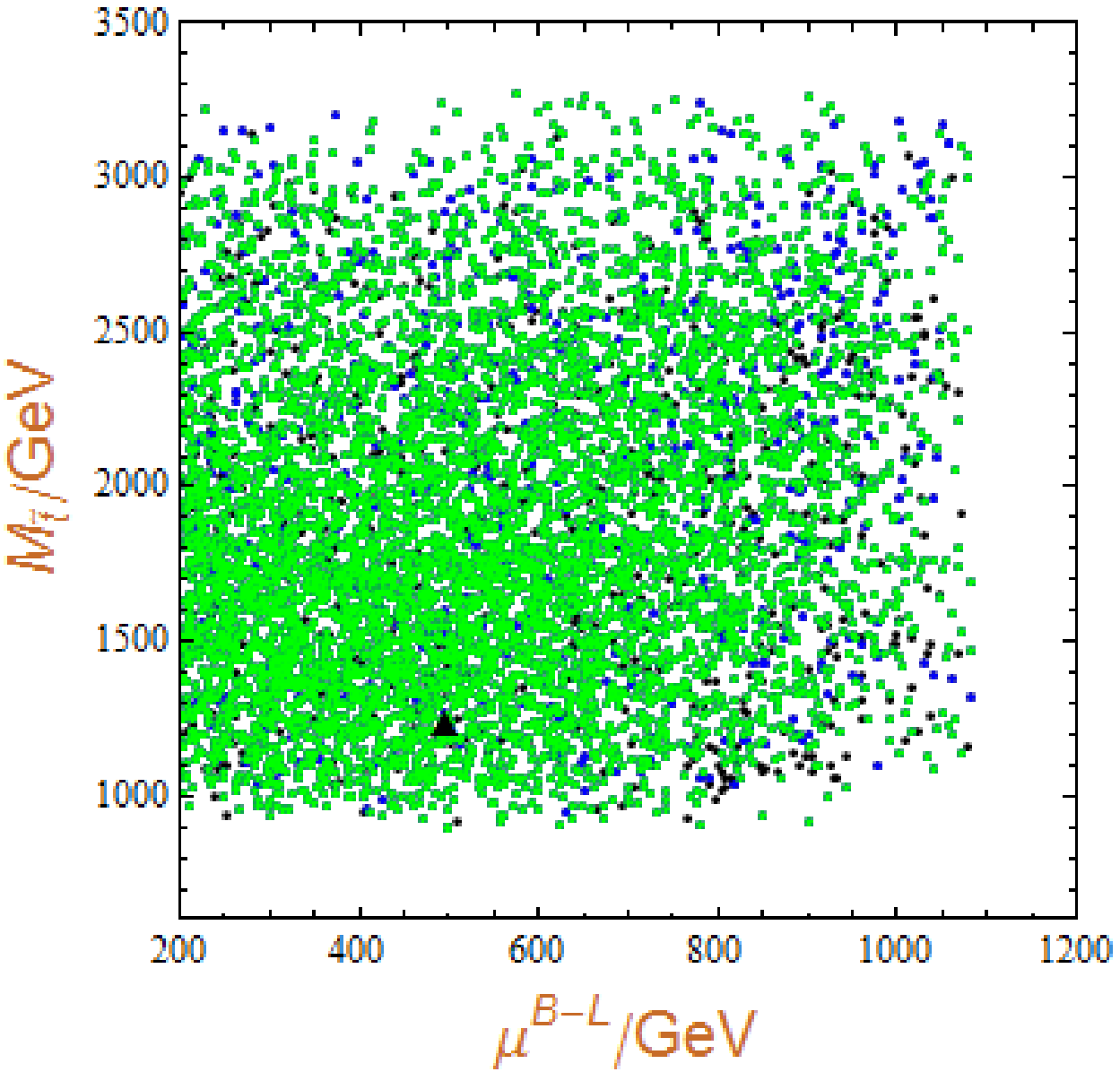}
\includegraphics[width=5cm]{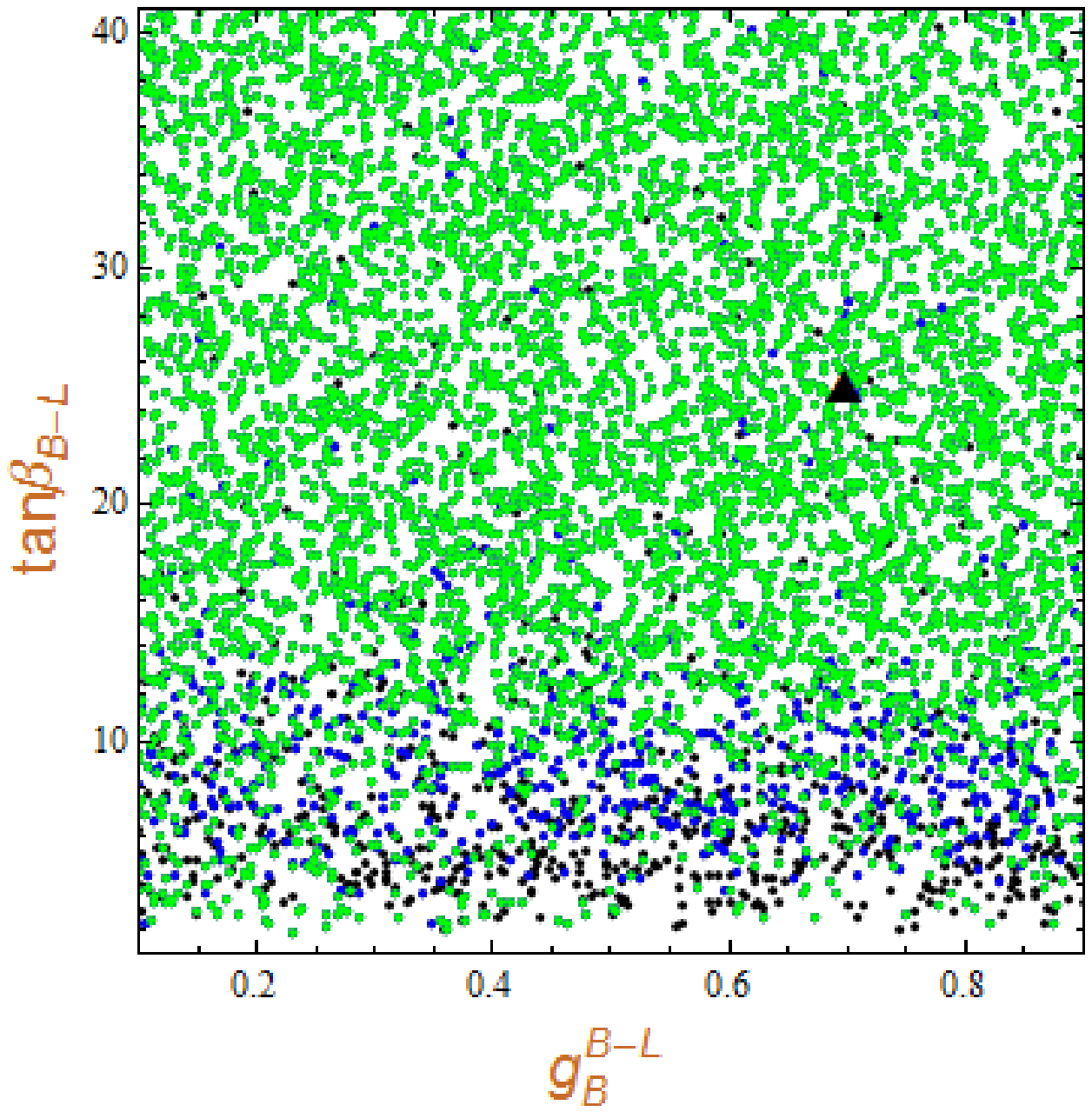}\\
\includegraphics[width=5.2cm]{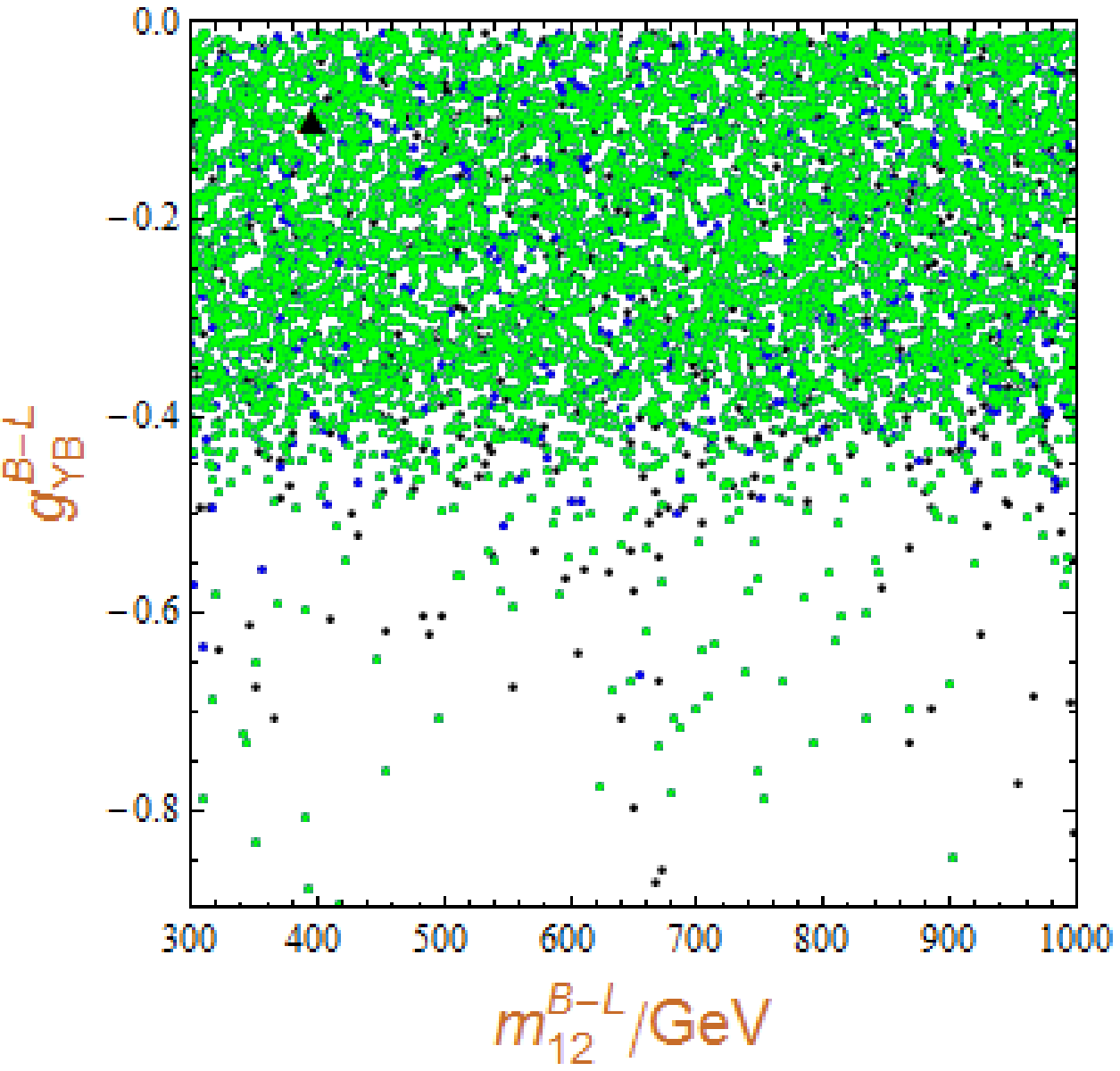}
\includegraphics[width=5.2cm]{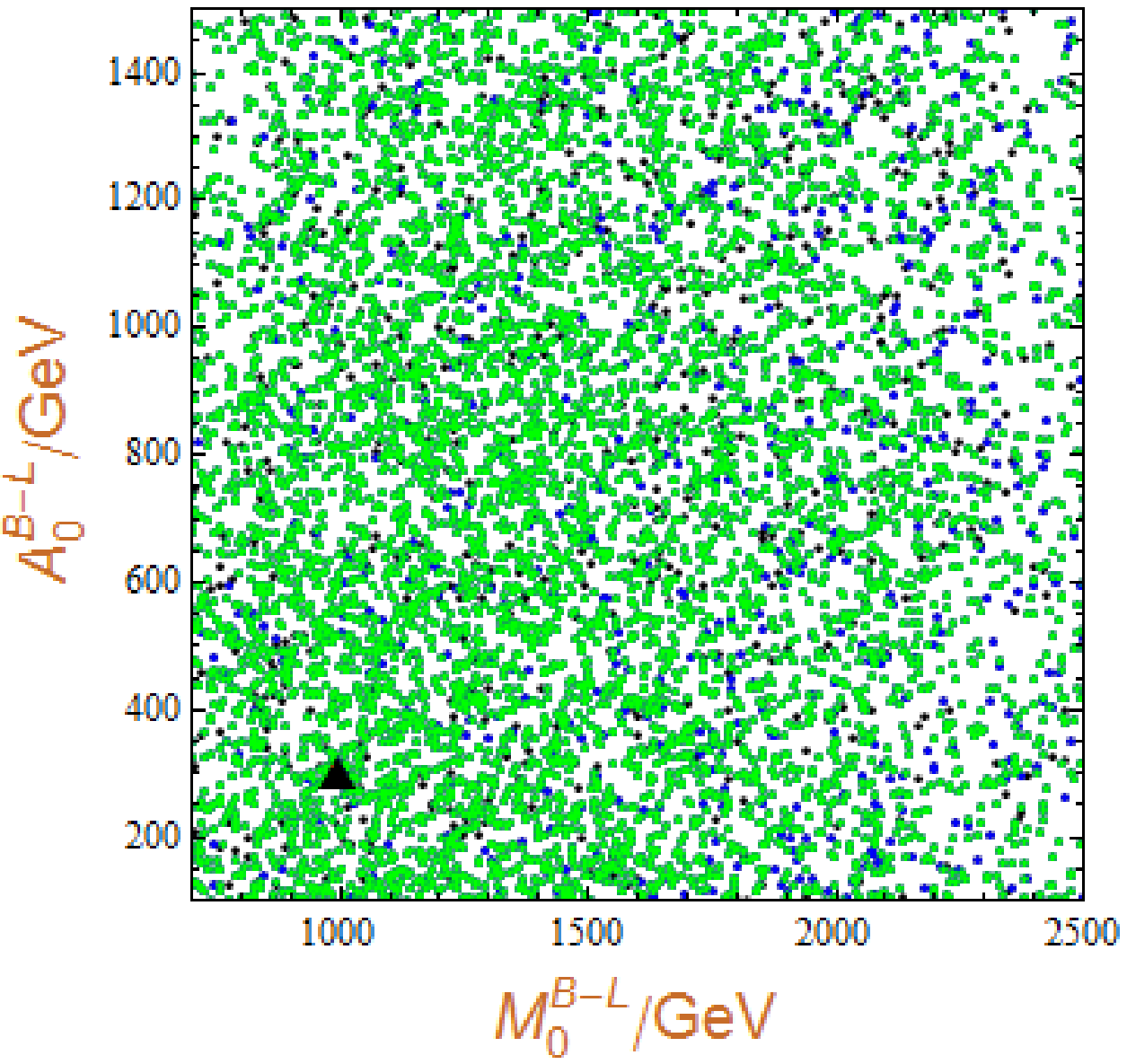}
\includegraphics[width=5cm]{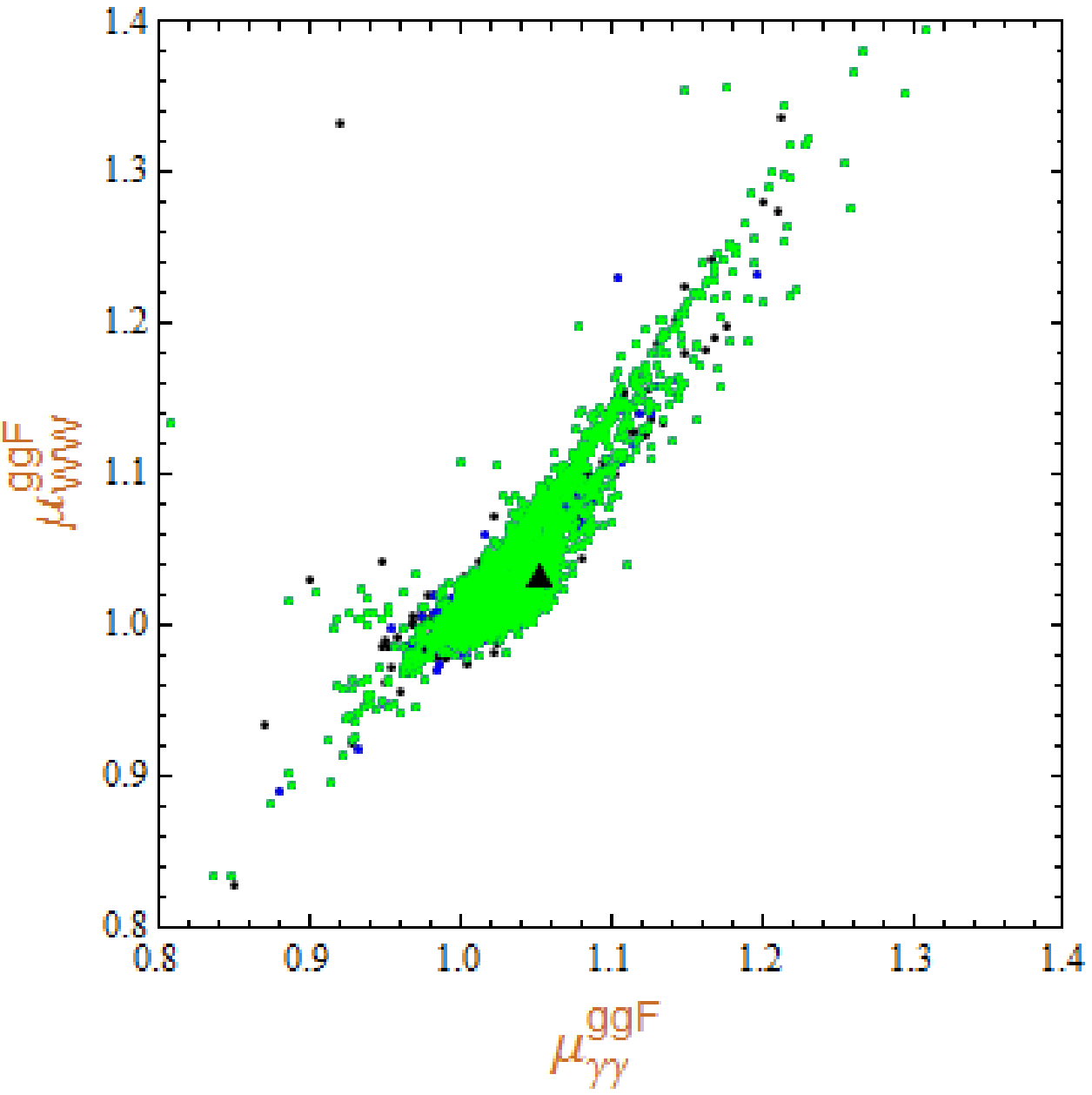}\\
\includegraphics[width=5.1cm]{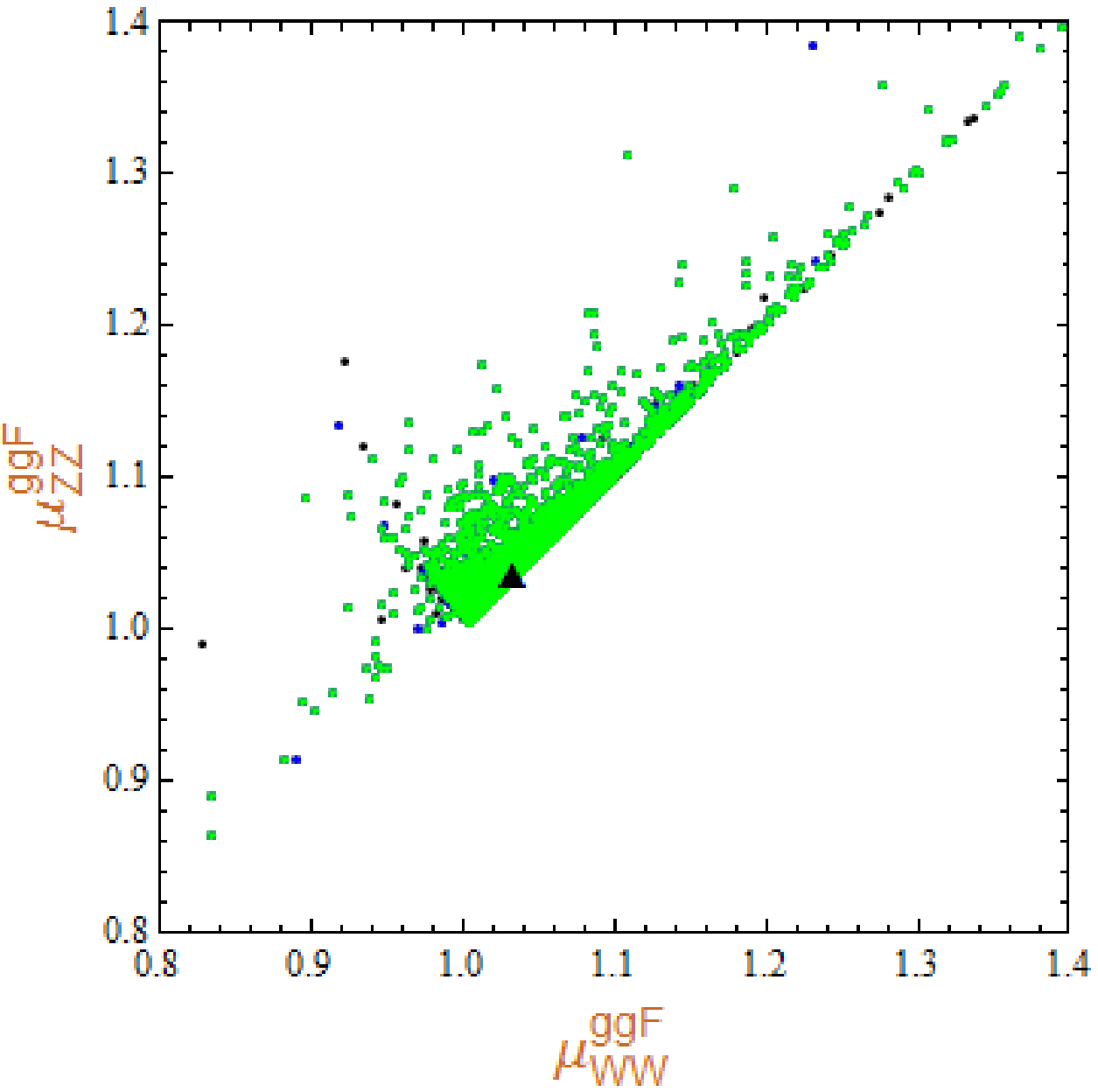}
\includegraphics[width=5.1cm]{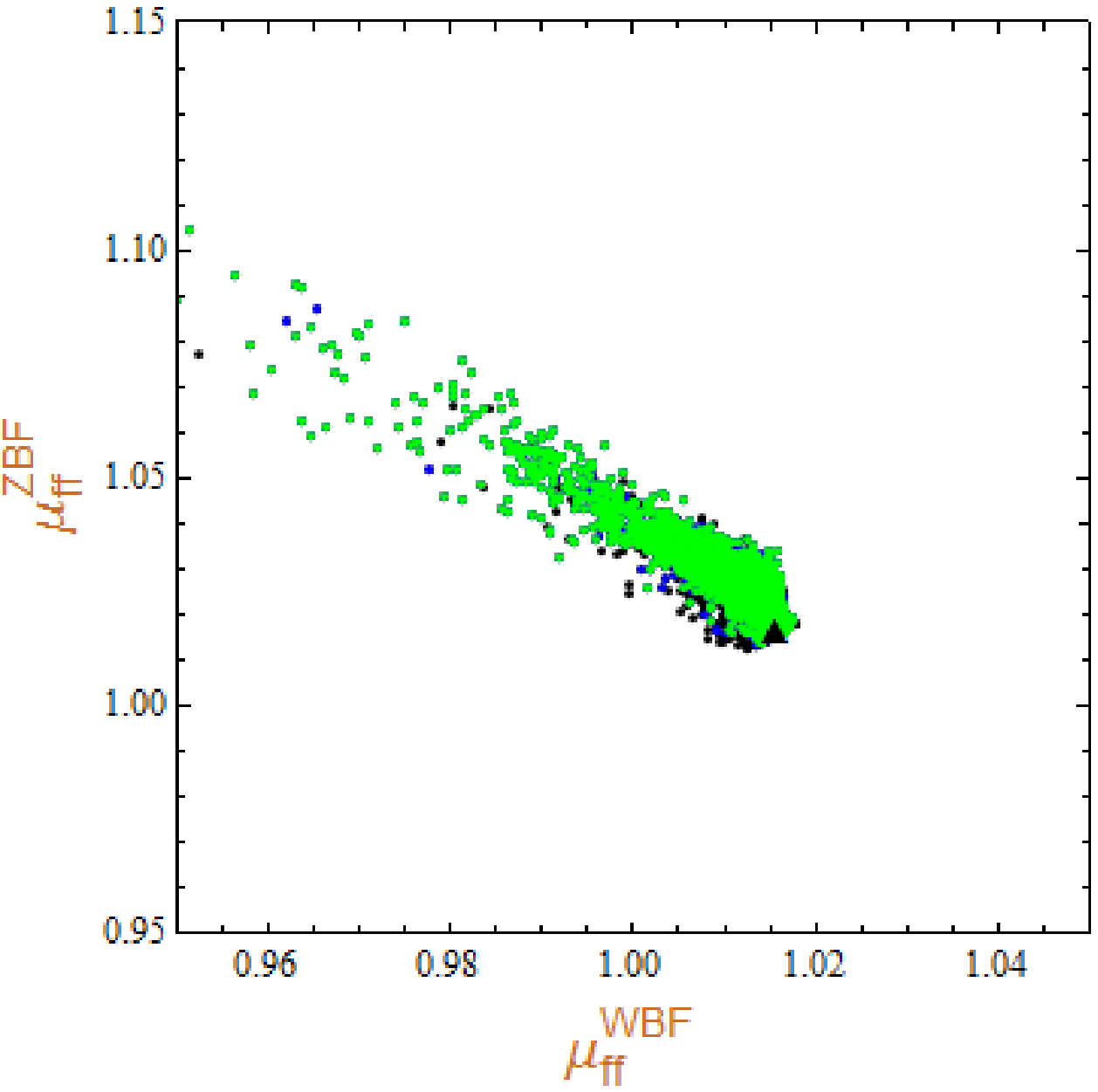}
\includegraphics[width=5.5cm]{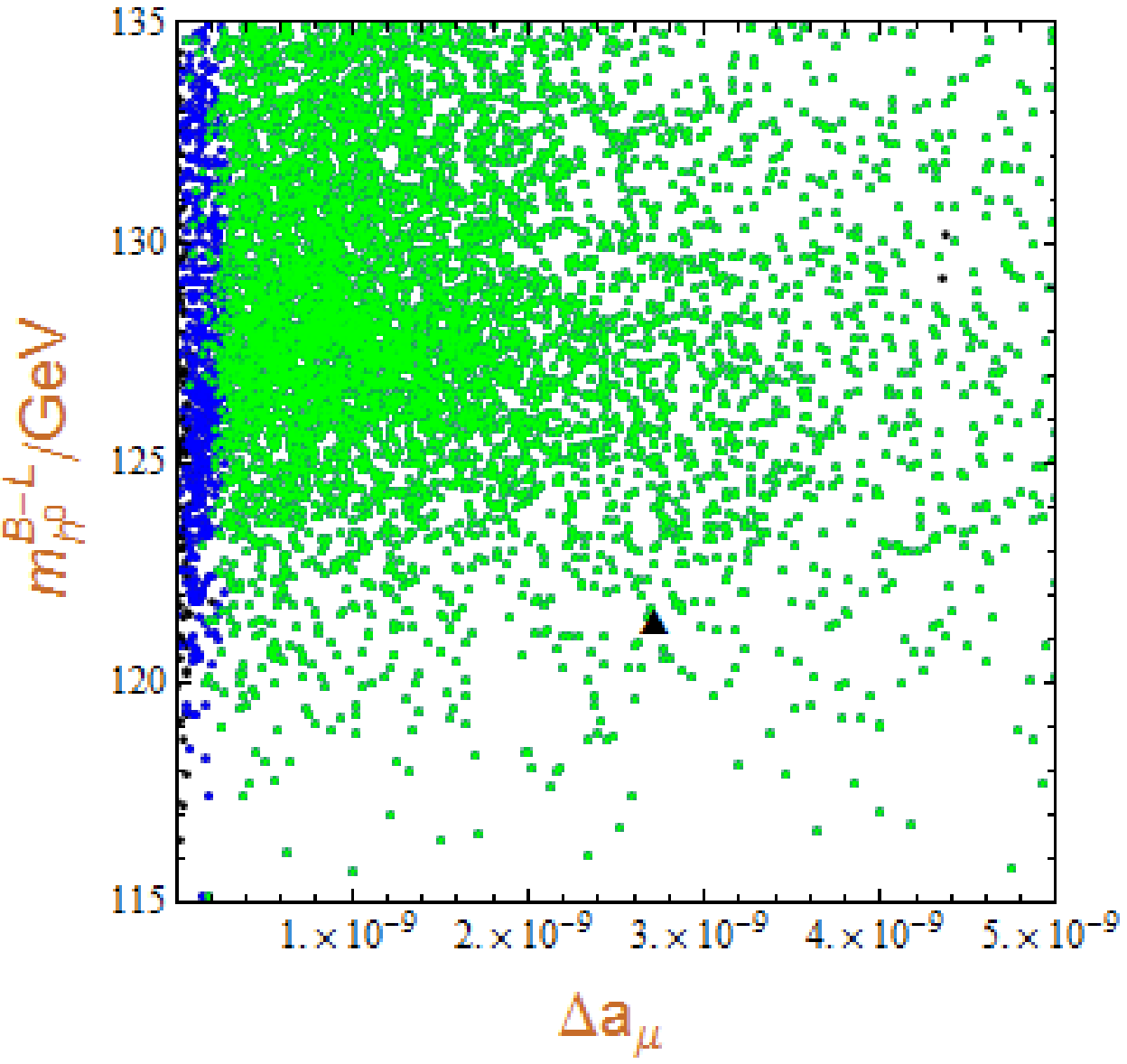}
\caption{The fitting results for B-LSSM with the $\chi^2$ analyses.} \label{fig3}
\end{figure}

Other than this, the B-LSSM numerical analyses are also taken over. We present the $\Delta_{FT}$ versus $\chi^2$, $\mu^{B-L}$ versus $M_{\tilde{t}}$, $g_{B}^{B-L}$ versus $\tan\beta_{B-L}$, $M_0^{B-L}$ versus $A_0^{B-L}$, $\mu_{\gamma\gamma}^{ggF}$ versus $\mu_{WW}^{ggF}$, $\mu_{WW}^{ggF}$ versus $\mu_{ZZ}^{ggF}$, $\mu_{ff}^{WBF}$ versus $\mu_{ff}^{ZBF}$, $\Delta a_{\mu}$ versus $m_{h^0}^{B-L}$ and $m_{12}^{B-L}$ wersus $g_{YB}^{B-L}$ in FIG.\ref{fig3} with $\Delta_{FT}$ around $20\sim150$. The black triangle shows the best-fitted benchmark point with minimal $(\chi^{B-L}_{min})^2$ = 2.47754. The green, blue, and black regions are respectively $90\%$, $95\%$, and $99\%$ confidence levels with $\chi^2 < (\chi^{B-L}_{min})^2+ 12.02$, $(\chi^{B-L}_{min})^2 + 14.07$ and $(\chi^{B-L}_{min})^2 + 18.49$. As fine-tuning fluctuates between $0.67\%$ and $5\%$, we observe that $0.2{\rm TeV}<\mu^{B-L}<1.0{\rm TeV}$, $1.0{\rm TeV}\lesssim M_{\tilde{t}}\lesssim 3.2{\rm TeV}$, $10\lesssim\tan\beta_{B-L}\lesssim40$ and $ -0.5\lesssim g_{YB}^{B-L}<0$ with $90\%$ confidence level. Under the above assumptions, the $\mu_{\gamma\gamma}^{ggF}$, $\mu_{WW}^{ggF}$ and $\mu_{ZZ}^{ggF}$ can be adjusted in the range of 1.0 to 1.2. The values of $\mu_{ff}^{WBF}$ and $\mu_{ff}^{ZBF}$ are approximately equal and tend to 1.0. And $\Delta a_{\mu}$ can be well corrected in the range of $0.5\times10^{-9}$ to $5.0\times10^{-9}$ for $\chi^2 < (\chi^{B-L}_{min})^2+ 12.02$, Therefore, all of aforementioned results are in good agreement with the corresponding experimental results.
\section{conclusion}
In this paper, we adopt the method of $\chi^2$ analyses in the BLMSSM and B-LSSM to calculate the Higgs mass, Higgs decays and muon $g-2$,  which will be better than MSSM. After scanning the parameter space, we point out some sensitive parameters in the BLMSSM and B-LSSM. In the BLMSSM, $g_{LB}^{BL}$ is changing from 0.1 to 0.9 while $\tan\beta_{BL}$ is limited in $4\sim40$ as the fine-tuning in the region $0.67\%-2.5\%$. As well as, we observe that $0.2{\rm TeV}<\mu^{B-L}<1.0{\rm TeV}$, $10\lesssim\tan\beta_{B-L}\lesssim40$ and $ -0.5\lesssim g_{YB}^{B-L}<0$ in the B-LSSM with fine-tuning fluctuating between $0.67\%$ and $5\%$. With the constraints $\mu\lesssim400\sqrt{\Delta_{FT}5\%}\;{\rm GeV}$, $M_{\tilde{t}}\lesssim1.2\sin\beta\sqrt{\Delta_{FT}5\%}\;{\rm TeV}$ and $115{\rm GeV}\lesssim m_{h^0}\lesssim135{\rm GeV}$, we can obtain the reasonable theoretical values for Higgs decays and muon $g-2$ respectively in the BLMSSM and B-LSSM, which are all in accordance with the experimental results. Other than this, the best-fitted benchmark points in the BLMSSM and B-LSSM will be acquired at minimal $(\chi^{BL}_{min})^2 = 2.34736$ and $(\chi^{B-L}_{min})^2 = 2.47754$, respectively. Therefore, the BLMSSM and B-LSSM are both more natural and realistic than MSSM.

{\bf Acknowledgments}

This work is supported by the Major Project of National Natural Science Foundation of China (NNSFC) (No. 11535002, No. 11705045), Post-graduate's Innovation Fund Project of Hebei Province with Grant No. CXZZBS2019027 and the youth top-notch talent support program of the Hebei Province.
\appendix
\section{the form factors}
The form factors are defined as
\begin{eqnarray}
&&A_{1/2}(x)=2\Big[x+(x-1)g(x)\Big]/x^2,\nonumber\\
&&A_0(x)=-(x-g(x))/x^2\;,\nonumber\\
&&A_1(x)=-\Big[2x^2+3x+3(2x-1)g(x)\Big]/x^2,\nonumber\\
&&g(x)=\left\{\begin{array}{l}\arcsin^2\sqrt{x},\;\;\;\;\;\;\;\;\;\;\;\;\;\;\;\;\;\;\;\;x\le1\\
-{1\over4}\Big[\ln{1+\sqrt{1-1/x}\over1-\sqrt{1-1/x}}-i\pi\Big]^2,\;x>1\;.\end{array}\right.
\label{g-function}
\end{eqnarray}
\begin{eqnarray}
&&F(x)=-(1-x^2)(\frac{47}{2}x^2-\frac{13}{2}+\frac{1}{x^2})-3(1-6x^2+4x^4)\ln x \nonumber \\&&\hspace{1.5cm}+\frac{3(1-8x^2+20x^4)}{\sqrt{4x^2-1}}\cos^{-1}\Big(\frac{3x^2-1}{2x^3}\Big)
\label{F-function}
\end{eqnarray}
 \begin{eqnarray}
\mathcal{B}(x,y)=\frac{1}{16 \pi
   ^2}\Big(\frac{x \ln x}{y-x}+\frac{y \ln
   y}{x-y}\Big),~~~
\mathcal{B}_1(x,y)=(
\frac{\partial }{\partial y}+\frac{y}{2}\frac{\partial^2 }{\partial y^2})\mathcal{B}(x,y).
\end{eqnarray}
\section{The expressions for Higgs decays in the B-LSSM}
The concrete expressions that present in the B-LSSM are specifically discussed in the following(in this part $i=1$):

1. CP-even Higgs-charge Higgs-charge Higgs contribution
\begin{eqnarray}
&&g^{B-L}_{h^0H^{\pm}H^{\pm}}=\frac{v}{2mz^2}\Big[\frac{1}{4} \Big(-2 g_{YB} g_{B}\Big(- v_{\bar{\eta}} Z_{{i 4}}^{H}  + v_{\eta} Z_{{i 3}}^{H} \Big)\Big(Z_{{j 1}}^{+} Z_{{k 1}}^{+}  - Z_{{j 2}}^{+} Z_{{k 2}}^{+} \Big)\nonumber \\&&\hspace{2cm}- Z_{{i 1}}^{H} \Big(Z_{{j 2}}^{+} \Big(- \Big(- g_{2}^{2}  + g_{1}^{2} + g_{Y B}^{2}\Big)v_d Z_{{k 2}}^{+}  + g_{2}^{2} v_u Z_{{k 1}}^{+} \Big)\nonumber \\&&
\hspace{2cm}+Z_{{j 1}}^{+} \Big(\Big(g_{1}^{2} + g_{Y B}^{2} + g_{2}^{2}\Big)v_d Z_{{k 1}}^{+}  + g_{2}^{2} v_u Z_{{k 2}}^{+} \Big)\Big)\nonumber \\
&&\hspace{2cm}+Z_{{i 2}}^{H} \Big(Z_{{j 1}}^{+} \Big(\Big(- g_{2}^{2}  + g_{1}^{2} + g_{Y B}^{2}\Big)v_u Z_{{k 1}}^{+}  - g_{2}^{2} v_d Z_{{k 2}}^{+} \Big)\nonumber \\
&&\hspace{2cm}- Z_{{j 2}}^{+} \Big(\Big(g_{1}^{2} + g_{Y B}^{2} + g_{2}^{2}\Big)v_u Z_{{k 2}}^{+}  + g_{2}^{2} v_d Z_{{k 1}}^{+} \Big)\Big)\Big)\Big];
\end{eqnarray}
2. CP-even Higgs-slepton-slepton contribution
\begin{eqnarray}
&&\hspace{-0.8cm}g^{B-L}_{h^0\tilde{L}\tilde{L}}=\frac{v}{2mz^2}\Big[\frac{1}{4} \Big(-2 \Big(\sqrt{2} \sum_{b=1}^{3}Z^{E,*}_{j b} \sum_{a=1}^{3}Z_{{k 3 + a}}^{E} T_{e,{a b}}   Z_{{i 1}}^{H} +\sqrt{2} \sum_{b=1}^{3}\sum_{a=1}^{3}Z^{E,*}_{j 3 + a} T^*_{e,{a b}}  Z_{{k b}}^{E}  Z_{{i 1}}^{H} \nonumber \\
&&+2 v_d \sum_{c=1}^{3}Z^{E,*}_{j 3 + c} \sum_{b=1}^{3}\sum_{a=1}^{3}Y^*_{e,{c a}} Y_{e,{b a}}  Z_{{k 3 + b}}^{E}   Z_{{i 1}}^{H} +2 v_d \sum_{c=1}^{3}\sum_{b=1}^{3}Z^{E,*}_{j b} \sum_{a=1}^{3}Y^*_{e,{a c}} Y_{e,{a b}}   Z_{{k c}}^{E}  Z_{{i 1}}^{H} \nonumber \\
&&- \sqrt{2} \mu^* \sum_{b=1}^{3}Z^{E,*}_{j b} \sum_{a=1}^{3}Y_{e,{a b}} Z_{{k 3 + a}}^{E}   Z_{{i 2}}^{H} - \sqrt{2} \mu \sum_{b=1}^{3}\sum_{a=1}^{3}Y^*_{e,{a b}} Z^{E,*}_{j 3 + a}  Z_{{k b}}^{E}  Z_{{i 2}}^{H} \Big)\nonumber \\
&&+\sum_{a=1}^{3}\hspace{-0.1cm}Z^{E,*}_{j 3 + a} Z_{{k 3 + a}}^{E}  \Big(\Big(2 g_{1}^{2} \hspace{-0.1cm}+\hspace{-0.1cm} g_{Y B} \Big(2 g_{Y B} \hspace{-0.1cm} +\hspace{-0.1cm} g_{B}\Big)\Big)v_d Z_{{i 1}}^{H} \hspace{-0.1cm}- \hspace{-0.1cm}\Big(2 g_{1}^{2} \hspace{-0.1cm}+\hspace{-0.1cm} g_{Y B} \Big(2 g_{Y B} \hspace{-0.1cm} +\hspace{-0.1cm} g_{B}\Big)\Big)v_u Z_{{i 2}}^{H} \nonumber \\
&&+2 \Big(2 g_{Y B} g_{B}  \hspace{-0.1cm}+ \hspace{-0.1cm}g_{B}^{2}\Big)\Big(\hspace{-0.1cm}-\hspace{-0.1cm} v_{\bar{\eta}} Z_{{i 4}}^{H} \hspace{-0.1cm} +\hspace{-0.1cm} v_{\eta} Z_{{i 3}}^{H} \Big)\hspace{-0.1cm}\Big)\hspace{-0.1cm}+\hspace{-0.1cm}\sum_{a=1}^{3}\hspace{-0.1cm}Z^{E,*}_{j a} Z_{{k a}}^{E}  \Big(\hspace{-0.1cm}-\hspace{-0.1cm} \Big(\hspace{-0.1cm}-\hspace{-0.1cm} g_{2}^{2} \hspace{-0.1cm} +\hspace{-0.1cm} g_{Y B} g_{B} \hspace{-0.1cm} +\hspace{-0.1cm} g_{1}^{2} \hspace{-0.1cm}+\hspace{-0.1cm} g_{Y B}^{2}\Big)v_d Z_{{i 1}}^{H}  \nonumber \\
&&+\Big(- g_{2}^{2}  + g_{Y B} g_{B}  + g_{1}^{2} + g_{Y B}^{2}\Big)v_u Z_{{i 2}}^{H}-2 \Big( g_{Y B} g_{B}  + g_{B}^{2}\Big)\Big(- v_{\bar{\eta}} Z_{{i 4}}^{H}  + v_{\eta} Z_{{i 3}}^{H} \Big)\Big)\Big)\Big];
\end{eqnarray}
3. CP-even Higgs-up squark-up squark contribution
\begin{eqnarray}
&&\hspace{-0.8cm}g^{B-L}_{h^0\tilde{U}\tilde{U}}=\frac{v}{2mz^2}\Big[\frac{1}{12} \delta_{\beta \gamma} \Big(6 \Big(\sqrt{2} \mu^* \sum_{b=1}^{3}Z^{U,*}_{j b}\hspace{-0.1cm} \sum_{a=1}^{3}\hspace{-0.1cm}Y_{u,{a b}} Z_{{k 3 + a}}^{U}   Z_{{i 1}}^{H} \hspace{-0.1cm}+\hspace{-0.1cm}\sqrt{2} \mu \sum_{b=1}^{3}\hspace{-0.1cm}\sum_{a=1}^{3}\hspace{-0.1cm}Y^*_{u,{a b}} Z^{U,*}_{j 3 + a}  Z_{{k b}}^{U}  Z_{{i 1}}^{H} \nonumber \\
&&- \Big(\sqrt{2} \sum_{b=1}^{3}Z^{U,*}_{j b} \sum_{a=1}^{3}Z_{{k 3 + a}}^{U} T_{u,{a b}}   +\sqrt{2} \sum_{b=1}^{3}\sum_{a=1}^{3}Z^{U,*}_{j 3 + a} T^*_{u,{a b}}  Z_{{k b}}^{U}  \nonumber \\
&&+2 v_u \Big(\sum_{c=1}^{3}Z^{U,*}_{j 3 + c} \sum_{b=1}^{3}\sum_{a=1}^{3}Y^*_{u,{c a}} Y_{u,{b a}}  Z_{{k 3 + b}}^{U}   + \sum_{c=1}^{3}\sum_{b=1}^{3}Z^{U,*}_{j b} \sum_{a=1}^{3}Y^*_{u,{a c}} Y_{u,{a b}}   Z_{{k c}}^{U} \Big)\Big)Z_{{i 2}}^{H} \Big)\nonumber \\
&&+\hspace{-0.1cm}\sum_{a=1}^{3}\hspace{-0.1cm}Z^{U,*}_{j 3 + a} Z_{{k 3 + a}}^{U}  \Big(\hspace{-0.1cm}-\hspace{-0.1cm} \Big(4 g_{1}^{2} \hspace{-0.1cm}+\hspace{-0.1cm} g_{Y B} \Big(4 g_{Y B} \hspace{-0.1cm} +\hspace{-0.1cm} g_{B}\Big)\Big)v_d Z_{{i 1}}^{H} \hspace{-0.1cm}+\hspace{-0.1cm}\Big(4 g_{1}^{2} \hspace{-0.1cm}+\hspace{-0.1cm} g_{Y B} \Big(4 g_{Y B} \hspace{-0.1cm} +\hspace{-0.1cm} g_{B}\Big)\Big)v_u Z_{{i 2}}^{H} \nonumber \\
&&-2 \Big( 4 g_{Y B} g_{B} \hspace{-0.1cm} +\hspace{-0.1cm} g_{B}^{2} \Big)\Big(\hspace{-0.1cm}-\hspace{-0.1cm} v_{\bar{\eta}} Z_{{i 4}}^{H} \hspace{-0.1cm} +\hspace{-0.1cm} v_{\eta} Z_{{i 3}}^{H} \Big)\Big)\hspace{-0.1cm}+\hspace{-0.1cm}\sum_{a=1}^{3}\hspace{-0.1cm}Z^{U,*}_{j a} Z_{{k a}}^{U}  \Big(\Big(\hspace{-0.1cm}-\hspace{-0.1cm}3 g_{2}^{2} \hspace{-0.1cm}+\hspace{-0.1cm} g_{Y B} g_{B} \hspace{-0.1cm} +\hspace{-0.1cm} g_{1}^{2}\hspace{-0.1cm} +\hspace{-0.1cm} g_{Y B}^{2}\Big)v_d Z_{{i 1}}^{H} \nonumber \\
&&- \Big(-3 g_{2}^{2} + g_{Y B} g_{B}  + g_{1}^{2} + g_{Y B}^{2}\Big)v_u Z_{{i 2}}^{H} +2 \Big(g_{Y B} g_{B}  + g_{B}^{2}\Big)\Big(- v_{\bar{\eta}} Z_{{i 4}}^{H}  + v_{\eta} Z_{{i 3}}^{H} \Big)\Big)\Big)\Big];
\end{eqnarray}
4. CP-even Higgs-down squark-down squark contribution
\begin{eqnarray}
&&\hspace{-0.8cm}g^{B-L}_{h^0\tilde{D}\tilde{D}}=\frac{v}{2mz^2}\Big[\frac{1}{12} \delta_{\beta \gamma} \Big(\hspace{-0.1cm}-\hspace{-0.1cm}6 \Big(\sqrt{2} \sum_{b=1}^{3}\hspace{-0.1cm}Z^{D,*}_{j b} \sum_{a=1}^{3}\hspace{-0.1cm}Z_{{k 3 + a}}^{D} T_{d,{a b}}   Z_{{i 1}}^{H} \hspace{-0.1cm}+\hspace{-0.1cm}\sqrt{2} \sum_{b=1}^{3}\hspace{-0.1cm}\sum_{a=1}^{3}\hspace{-0.1cm}Z^{D,*}_{j 3 + a} T^*_{d,{a b}}  Z_{{k b}}^{D}  Z_{{i 1}}^{H} \nonumber \\
&&+2 v_d \sum_{c=1}^{3}Z^{D,*}_{j 3 + c} \sum_{b=1}^{3}\sum_{a=1}^{3}Y^*_{d,{c a}} Y_{d,{b a}}  Z_{{k 3 + b}}^{D}   Z_{{i 1}}^{H} +2 v_d \sum_{c=1}^{3}\sum_{b=1}^{3}Z^{D,*}_{j b} \sum_{a=1}^{3}Y^*_{d,{a c}} Y_{d,{a b}}   Z_{{k c}}^{D}  Z_{{i 1}}^{H} \nonumber \\
&&- \sqrt{2} \mu^* \sum_{b=1}^{3}Z^{D,*}_{j b} \sum_{a=1}^{3}Y_{d,{a b}} Z_{{k 3 + a}}^{D}   Z_{{i 2}}^{H} - \sqrt{2} \mu \sum_{b=1}^{3}\sum_{a=1}^{3}Y^*_{d,{a b}} Z^{D,*}_{j 3 + a}  Z_{{k b}}^{D}  Z_{{i 2}}^{H} \Big)\nonumber \\
&&+\hspace{-0.1cm}\sum_{a=1}^{3}\hspace{-0.1cm}Z^{D,*}_{j 3 + a} Z_{{k 3 + a}}^{D}  \Big(\Big(2 g_{1}^{2} \hspace{-0.1cm}+\hspace{-0.1cm} g_{Y B} \Big(2 g_{Y B}  \hspace{-0.1cm}-\hspace{-0.1cm} g_{B} \Big)\Big)v_d Z_{{i 1}}^{H} \hspace{-0.1cm}+\hspace{-0.1cm}\Big(\hspace{-0.1cm}-\hspace{-0.1cm}2 g_{1}^{2} \hspace{-0.1cm}+\hspace{-0.1cm} g_{Y B} \Big(\hspace{-0.1cm}-\hspace{-0.1cm}2 g_{Y B} \hspace{-0.1cm} + \hspace{-0.1cm} g_{B}\Big)\Big)v_u Z_{{i 2}}^{H} \nonumber \\
&&+2 \Big(2 g_{Y B} g_{B}  \hspace{-0.1cm}-\hspace{-0.1cm} g_{B}^{2}\Big)\Big(\hspace{-0.1cm}- \hspace{-0.1cm} v_{\bar{\eta}} Z_{{i 4}}^{H} \hspace{-0.1cm} +\hspace{-0.1cm} v_{\eta} Z_{{i 3}}^{H} \Big)\Big)\hspace{-0.1cm}+\hspace{-0.1cm}\sum_{a=1}^{3}\hspace{-0.1cm}Z^{D,*}_{j a} Z_{{k a}}^{D}  \Big(\Big(3 g_{2}^{2} \hspace{-0.1cm}+\hspace{-0.1cm} g_{Y B} g_{B} \hspace{-0.1cm} +\hspace{-0.1cm} g_{1}^{2}\hspace{-0.1cm} +\hspace{-0.1cm} g_{Y B}^{2}\Big)v_d Z_{{i 1}}^{H} \nonumber \\
&&- \Big(3 g_{2}^{2} + g_{Y B} g_{B}  + g_{1}^{2} + g_{Y B}^{2}\Big)v_u Z_{{i 2}}^{H} +2 \Big( g_{Y B} g_{B}  + g_{B}^{2}\Big)\Big(- v_{\bar{\eta}} Z_{{i 4}}^{H}  + v_{\eta} Z_{{i 3}}^{H} \Big)\Big)\Big)\Big];
\end{eqnarray}
5. CP-even Higgs-W boson-W boson contribution
\begin{eqnarray}
&&g^{B-L}_{h^0WW}=\Big(\cos\beta Z_{{i 1}}^{H}  + \sin\beta Z_{{i 2}}^{H} \Big);
\end{eqnarray}
6. CP-even Higgs-Z boson-Z boson contribution
\begin{eqnarray}
&&\hspace{-0.5cm}g^{B-L}_{h^0ZZ}=\frac{v}{2mz^2}\Big[\frac{1}{2} \Big(v_d \Big(g_1 \cos{\Theta'}_W  \sin\Theta_W   + g_2 \cos\Theta_W  \cos{\Theta'}_W  \hspace{-0.1cm} - \hspace{-0.1cm}g_{Y B} \sin{\Theta'}_W  \Big)^{2} Z_{{i 1}}^{H} \nonumber \\
&&\hspace{1cm}+v_u \Big(g_1 \cos{\Theta'}_W  \sin\Theta_W  \hspace{-0.1cm} +\hspace{-0.1cm} g_2 \cos\Theta_W  \cos{\Theta'}_W  \hspace{-0.1cm} -\hspace{-0.1cm} g_{Y B} \sin{\Theta'}_W  \Big)^{2} Z_{{i 2}}^{H} \nonumber \\
&&\hspace{1cm}+4 \Big(- g_{B} \sin{\Theta'}_W \Big)^{2} \Big(v_{\bar{\eta}} Z_{{i 4}}^{H}  + v_{\eta} Z_{{i 3}}^{H} \Big)\Big)\Big(g_{\sigma \mu}\Big)\Big];
\end{eqnarray}
7. CP-even Higgs-chargino-chargino contribution
\begin{eqnarray}
&&g^{B-L}_{h^0\chi^{\pm}\chi^{\pm}}=-\frac{v}{m_{\chi^{\pm}}}\Big[-\frac{1}{\sqrt{2}} g_2 \Big(U_{{k 1}} V_{{j 2}} Z_{{i 2}}^{H}  + U_{{k 2}} V_{{j 1}} Z_{{i 1}}^{H} \Big)];
\end{eqnarray}
8. CP-even Higgs-down quark-down quark contribution
\begin{eqnarray}
&&g^{B-L}_{h^0 d d}=-\frac{v}{m_{d}}\Big[-\frac{1}{\sqrt{2}} \delta_{\alpha \beta} \sum_{b=1}^{3}\sum_{a=1}^{3}Y^*_{d,{a b}} U_{R,{j a}}^{d}  U_{L,{k b}}^{d}  Z_{{i 1}}^{H} \Big];
\end{eqnarray}
9. CP-even Higgs-up quark-up quark contribution
\begin{eqnarray}
&&g^{B-L}_{h^0 u u}=-\frac{v}{m_{u}}\Big[-\frac{1}{\sqrt{2}} \delta_{\alpha \beta} \sum_{b=1}^{3}\sum_{a=1}^{3}Y^*_{u,{a b}} U_{R,{j a}}^{u}  U_{L,{k b}}^{u}  Z_{{i 2}}^{H} \Big];
\end{eqnarray}
10. CP-even Higgs-lepton-lepton contribution
\begin{eqnarray}
&&g^{B-L}_{h^0 l l}=-\frac{v}{m_{l}}\Big[-\frac{1}{\sqrt{2}} \sum_{b=1}^{3}\sum_{a=1}^{3}Y^*_{e,{a b}} U_{R,{j a}}^{e}  U_{L,{k b}}^{e}  Z_{{i 1}}^{H} \Big].
\end{eqnarray}
\section{The expressions for muon $(g-2)$ in the B-LSSM}
In the B-LSSM, the one loop corrections for $(g-2)_{\mu}$ are mainly affected by slepton-neutralino, CP-odd sneutrino-chargino and CP-even sneutrino-chargino contributions. The concrete expressions for $(\mathcal{A}_1)^I,(\mathcal{A}_2)^I,(\mathcal{C}_1)^I$ and $(\mathcal{C}_2)^I$ that present in the B-LSSM can be specifically discussed as
\begin{eqnarray}
&&\hspace{-1cm}(\mathcal{A}_1^{B-L})^I_{\tilde{L}\chi^0}\hspace{-0.1cm}=\hspace{-0.1cm}\frac{1}{2}\Big[\sqrt{2}g_1N_{i1}^*\sum_{a=1}^3U_{L,ja}^{e*}Z_{ka}^E
\hspace{-0.1cm}+\hspace{-0.1cm}\sqrt{2}g_2N_{i2}^*\sum_{a=1}^3U_{L,ja}^{e*}Z_{ka}^E\hspace{-0.1cm}+\hspace{-0.1cm}\sqrt{2}g_{YB}N_{i5}^*\sum_{a=1}^3U_{L,ja}^{e*}Z_{ka}^E
\nonumber\\&&\hspace{2.5cm}\hspace{-0.1cm}+\hspace{-0.1cm}\sqrt{2}g_BN_{i5}^*\sum_{a=1}^3U_{L,ja}^{e*}Z_{ka}^E
\hspace{-0.1cm}-\hspace{-0.1cm}2N_{i3}^*\sum_{b=1}^3U_{L,jb}^{e*}\sum_{a=1}^3Y_{e,ab}Z_{k,3+a}^E\Big];\nonumber \\&&\hspace{-1cm}(\mathcal{A}_2^{B-L})^I_{\tilde{L}\chi^0}\hspace{-0.1cm}=\hspace{-0.1cm}-\frac{1}{\sqrt{2}}\Big[\hspace{-0.1cm}\sum_{a=1}^3\hspace{-0.1cm}Z_{k,3+a}^EU_{R,ja}^{e}
(2g_1N_{i1}\hspace{-0.1cm}+\hspace{-0.1cm}(2g_{YB}\hspace{-0.1cm}+\hspace{-0.1cm}g_B)N_{i5})\hspace{-0.1cm}+\hspace{-0.1cm}\sum_{b=1}^3\sum_{a=1}^3\hspace{-0.1cm}Y_{e,ab}^*U_{R,ja}^{e}Z_{kb}^EN_{i3}\Big];
\end{eqnarray}
\begin{eqnarray}
&&(\mathcal{C}_1^{B-L})^I_{\chi^-\tilde{\nu}^i}=\frac{i}{\sqrt{2}}U_{j2}^*\sum_{b=1}^3Z_{kb}^{i,*}\sum_{a=1}^3U_{R,ia}^{e,*}Y_{e,ab};\nonumber \\&&(\mathcal{C}_2^{B-L})^I_{\chi^-\tilde{\nu}^i}=-\frac{i}{\sqrt{2}}\Big[g_2\sum_{a=1}^3Z_{ka}^{i,*}U_{L,ia}^{e}V_{j1}-
\sum_{b=1}^3\sum_{a=1}^3Y_{\nu,ab}^*Z_{k,3+a}^{i,*}U_{L,ib}^eV_{j2}\Big];
\end{eqnarray}
\begin{eqnarray}
&&(\mathcal{C}_1^{B-L})^I_{\chi^-\tilde{\nu}^R}=\frac{1}{\sqrt{2}}U_{j2}^*\sum_{b=1}^3Z_{kb}^{R,*}\sum_{a=1}^3U_{R,ia}^{e,*}Y_{e,ab};\nonumber \\&&(\mathcal{C}_2^{B-L})^I_{\chi^-\tilde{\nu}^R}=\frac{1}{\sqrt{2}}\Big[-g_2\sum_{a=1}^3Z_{ka}^{R,*}U_{L,ia}^{e}V_{j1}+
\sum_{b=1}^3\sum_{a=1}^3Y_{\nu,ab}^*Z_{k,3+a}^{R,*}U_{L,ib}^eV_{j2}\Big].
\end{eqnarray}

 \end{document}